\documentclass[prd,superscriptaddress,nofootinbib,preprintnumbers,amsmath,amssymb,10pt,twocolumn]{revtex4-1}
\usepackage[utf8]{inputenc}
\usepackage{amsfonts}
\usepackage{mathrsfs}
\usepackage{physics}
\usepackage{footnote}
\usepackage{scrextend}
\usepackage{leftidx}
\usepackage{soul}
\usepackage{braket}
\usepackage{makecell}
 \usepackage{multirow}
 \usepackage{array}
\usepackage{comment}
\usepackage{tabularx}
\usepackage{slashed}
\usepackage{listings}
\usepackage[lofdepth,lotdepth,caption=false]{subfig}
 \newcolumntype{C}[1]{>{\centering\arraybackslash}p{#1}}
 
\usepackage[dvipsnames]{xcolor}
\definecolor{red}{rgb}{0.9, 0,0}
\definecolor{cerulean}{rgb}{0., 0.42,0.9}
\definecolor{richlavender}{rgb}{0.67, 0.38, 0.8}

\usepackage{epsfig}
\usepackage{graphicx}  
\usepackage{url}
\usepackage{color}
\usepackage{hyperref}
\hypersetup{
     colorlinks   = true,
     citecolor    = red,
	 linkcolor=blue
}
\usepackage{float}
\usepackage[export]{adjustbox}

\usepackage[capitalise]{cleveref}

\renewcommand{\ket}{\rangle}

\definecolor{red}{rgb}{0.9, 0,0}
\definecolor{cerulean}{rgb}{0., 0.62,0.9}
\definecolor{navy}{rgb}{0.05, 0.05,0.8}

\graphicspath{{./figures/}}

\newcommand{\bfq}{{\bf q}}

\newcommand{\bfr}{{\bf r}}

\newcommand{\bfv}{{\bf v}}
\newcommand{\bfJ}{{\bf J}}

\usepackage[dvipsnames]{xcolor}

\makeatletter
\def\l@subsection#1#2{}
\def\l@subsubsection#1#2{}
\makeatother

\begin{document}

\title{
Spin-Dependent Scattering of Sub-GeV Dark Matter: Models and Constraints
}

\author{Stefania Gori}
\affiliation{Department of Physics and Santa Cruz Institute for Particle Physics, University of California Santa Cruz, Santa Cruz, CA 95064, USA}
\author{Simon Knapen}
\affiliation{Theoretical Physics Group, Lawrence Berkeley National Laboratory, Berkeley, CA 94720, USA}
\affiliation{Berkeley Center for Theoretical Physics, Department of Physics, University of California, Berkeley, CA 94720, USA}
\author{Tongyan Lin}
\affiliation{ Department of Physics, University of California, San Diego, CA 92093, USA }
\author{Pankaj Munbodh}
\affiliation{Department of Physics and Santa Cruz Institute for Particle Physics, University of California Santa Cruz, Santa Cruz, CA 95064, USA}
\author{Bethany Suter}
\affiliation{Theoretical Physics Group, Lawrence Berkeley National Laboratory, Berkeley, CA 94720, USA}
\affiliation{Berkeley Center for Theoretical Physics, Department of Physics, University of California, Berkeley, CA 94720, USA}

\date{\today}

\begin{abstract}\noindent 

We calculate the scattering rate of sub-GeV dark matter in solid-state targets for spin-dependent dark matter -- nucleon interactions.  For dark matter particles with mass below 100 MeV, the scattering occurs predominantly through incoherent phonon production. 
For dark matter heavier than 100 MeV, we match onto the nuclear recoil calculation.
To compare the sensitivity of future direct detection experiments with existing constraints, we consider three models with interactions which are mediated by spin-0 or spin-1 particles.
This allows us to derive bounds on the cross section from searches for the mediating particle, including bounds from stellar cooling, beam dump experiments, meson factories and dark matter self-interactions. 
The existing bounds are very stringent, though for $m_\chi\gtrsim 100$ MeV there is  parameter space which may be accessible with direct detection, depending on the exposure and background rates.

\end{abstract}

\maketitle


\renewcommand{\baselinestretch}{0.15}\normalsize
\tableofcontents
\renewcommand{\baselinestretch}{1.0}\normalsize


\section{Introduction}

Advancements in experimental techniques, enabling increasingly lower energy thresholds, have been pushing direct detection of dark matter (DM) into the sub-GeV mass range \cite{Essig:2022dfa}. 
For DM which primarily couples to hadronic matter, the main signature in a solid or liquid target is a very soft nuclear recoil, potentially generating just a few phonons depending on the DM mass.
The TESSERACT collaboration recently released their first science run for this signature, with a sub-eV energy resolution \cite{TESSERACT:2025tfw}.
This progress necessitates complementary theoretical work on two fronts. First, to achieve detectable scattering cross sections, a relatively low mass mediator particle is needed. 
Existing constraints on such mediators must be then translated into upper bounds on the direct detection rate accessible to future experiments.
Second, for sub-GeV DM, the momentum transfer can be sufficiently small that the standard free elastic nuclear recoil approximation breaks down entirely.
In this case, DM scattering excites collective modes with wavelengths longer or comparable to the interatomic spacing in the target material. 

Both questions have been comprehensively addressed for models in which the dark matter couples to nuclei primarily through a spin-independent interaction.
Concretely, strong upper bounds on the direct detection cross section were obtained by combining bounds from stellar cooling, meson factories, cosmology and galaxy halo shapes, see e.g.~\cite{Knapen:2017xzo,Green:2017ybv,Batell:2018fqo,Dvorkin:2019zdi,Bondarenko:2019vrb,Chang:2019xva,Elor:2021swj,Balan:2024cmq,Cox:2024rew}.  
These bounds indicate that the most viable parameter space exists for models with a light dark photon mediator, or models where the particle being detected is only a subcomponent of the full dark matter density.
In addition, extensive rate calculations have been performed for a wide range of materials, for scattering into both single \cite{Knapen:2017ekk,Griffin:2018bjn,Cox:2019cod,Trickle:2019nya,Griffin:2019mvc,Caputo:2019xum,Griffin:2020lgd,Trickle:2020oki,Taufertshofer:2023rgq,Ashour:2024xfp} and multiple phonons \cite{Knapen:2016cue,Campbell-Deem:2019hdx,Kahn:2020fef,Campbell-Deem:2022fqm,Caputo:2020sys,Lin:2023slv,Stratman:2024sng}. 
For anisotropic materials, the daily modulation of the scattering rate has also been computed \cite{Griffin:2018bjn,Coskuner:2021qxo,Griffin:2020lgd,Taufertshofer:2023rgq,Stratman:2024sng}. 
For recent reviews on this subject and other detection channels, see~\cite{Kahn:2021ttr,Zurek:2024qfm}.

In this work, we consider models where the dark matter primarily has \emph{spin-dependent} interactions with nuclei. We calculate the scattering rate in crystals\footnote{Spin-dependent interactions have also been considered in the context of molecules \cite{Essig:2019kfe}. }, and survey the existing constraints on the cross section.
We start in \cref{sec:modelsandconstraints} by defining three benchmark models with spin-0 and spin-1 mediators, which generate the most motivated, non-relativistic operators for spin-dependent scattering. 
For each model, we must define a UV-completion in order to determine the bounds on the mediator particle and scattering rate. Since our goal is to find the maximum available parameter space for direct detection, we make choices which \emph{minimizes the strength of the existing bounds}, while avoiding severe fine-tuning and/or elaborate dark sector model building. While it is inevitable that all bounds are model-dependent to a degree, we claim that alternative model building choices would generally strengthen the bounds. This section provides a brief summary of our choice of UV-completion, and establishes notation and conventions for the models, such that readers primarily interested in the results can directly proceed to \cref{sec:discussion}.

In \cref{sec:scalarbounds,sec:modelAp} we provide more details on the models for spin-0 and spin-1 (axial vector) mediators, respectively. We survey the latest bounds in each case.
The coupling of the mediator to the Standard Model (SM) is primarily constrained by stellar cooling, beam dump experiments, meson factories, and LHC searches.  
The coupling of the dark matter with the mediator is constrained by dark matter self-interaction bounds and/or theoretical consistency requirements.
Combining these bounds yields an upper bound on the allowed direct detection cross section (see~\cite{Ramani:2019jam} for an earlier study).

~\cref{sec:scatteringcalculation} describes the calculation for spin-dependent scattering into a (multi-)phonon final state.
The scattering rate depends qualitatively on whether the nuclear spins in the crystal are aligned. If the nuclear spins have a high degree of coherence across multiple crystal cells, then coherent scattering contributes, similar to the case of spin-independent scattering (for a general discussion of the formalism for spin-dependent interactions in this case, see \cite{Trickle:2020oki}).
While spin-polarized targets have been considered in the context of nuclear recoils \cite{Franarin:2016ppr,Catena:2018uae,Jenks:2022wtj}, they likely pose additional experimental challenges. 
We therefore consider crystals in which the nuclear spins are \emph{oriented randomly}, such that scattering is fully incoherent.
Our calculation bears some similarities with the multiphonon computations for spin-independent interactions at moderate momentum transfer ($q \sim 10-100$ keV), where the incoherent rate can be a good approximation to the coherent rate \cite{Campbell-Deem:2022fqm}. 
Incoherent phonon production in unpolarized crystals was also considered in the context of non-resonant axion absorption \cite{Bloch:2024qqo}.
We have implemented the full calculation in the public code \textbf{DarkELF} \cite{Knapen:2021bwg}, as documented in \cref{app:darkelf}.

Our main results can be found in \cref{sec:discussion}. 
We compare the upper bounds on spin-dependent direct detection cross sections with the sensitivity of an idealized  future experiment. 
Our qualitative conclusions are summarized in \cref{sec:summary}. The appendices contain additional results, as well as details on the UV completions and the \textbf{DarkELF} implementation of our calculations.

\section{Models and constraints\label{sec:modelsandconstraints}}

In order to obtain scattering rates which are large enough to be relevant for a near-future direct detection experiment, we must  assume the existence of a new mediator with mass comparable to or lower than the dark matter.
To reliably map the parameter space available to direct detection, we must therefore account for the existing bounds on such mediators.
This can only be done by committing to one or more benchmark models, as we do in this section. 




We consider three models which generate spin-dependent couplings to nuclei. As effective theories, they are defined by 
\begin{align}
\label{eq:UV_axion}
    \mathcal{L}_{\phi} &= \phi\left[g_\chi  \bar{\chi} \chi + g_p  \bar{p}\gamma^5 p + g_n  \bar{n} \gamma^5 n\right]\\
\label{eq:UV_axion2}
    \mathcal{L}_{a} &= a\left[ g_\chi  \bar{\chi}\gamma_5 \chi + g_p  \bar{p}\gamma^5 p + g_n  \bar{n} \gamma^5 n\right]\\
\label{eq:UV_vector}
    \mathcal{L}_{A'} &= A'^\mu\left[g_\chi  \bar{\chi}\gamma_\mu\gamma_5 \chi + g_p  \bar{p}\gamma_\mu\gamma_5 p + g_n \bar{n} \gamma_\mu\gamma_5 n\right],
\end{align}
with $\chi$ the dark matter, which we take to be a Dirac fermion. $n$ and $p$ are the SM neutron and proton, respectively. The mediators $\phi,a$ and $A'$ couple to the SM as a pseudoscalar and pseudovector, respectively. 

The most relevant bounds on these interactions depend on the UV completion. In the remainder of this section, we explain our choices for the UV completion and provide a brief summary of the bounds  (see also \cref{tab:boundssummary}). We aim for this section to provide all the relevant notation and main assumptions, such that readers interested in the results can directly go to \cref{sec:discussion}. A more detailed discussion of the bounds will be provided in \cref{sec:scalarbounds} for $\phi$ and $a$ mediators and \cref{sec:modelAp} for the $A'$. 

{\boldmath\bf{Spin-0 mediators $\phi$ and $a$:}} The possible UV completions of the $g_{p,n}$ effective interaction are well known in the context of axion-like particles (ALPs). Here we consider the operators $\phi G \tilde G$ and $a G \tilde G$ as the origin of the nucleon interaction, as the constraints are generally weaker than coupling $\phi/a$ universally to the quarks. The most stringent bounds on $g_p$ and $g_n$ then come from exotic meson decays and beam-dump experiments for \mbox{$ m_{a,\phi} \gtrsim 300$ MeV}, and from supernova SN1987A for \mbox{$ m_{a,\phi} \lesssim 300$ MeV}. Additionally, for the $\phi$ mediator, there are stringent bounds on $g_\chi$ from dark matter self-interactions, whereas for the $a$ mediator, self-interaction rates have an additional velocity suppression so the self-interacting dark matter (SIDM) bounds on $g_\chi$ are weaker. In this case, we will show that for $m_a \gtrsim 0.1$ MeV the only constraint on $g_\chi$ comes from unitarity.

{\boldmath\bf{Axial vector mediator $A'$:}} This case is more subtle, as it requires a new $U(1)^\prime$ gauge symmetry which is anomalous in the infrared.
We construct an anomaly-free UV completion in \cref{app:AxialVectorModel} by introducing a set of heavy colored fermions which are chiral with respect to the $U(1)^\prime$.
We furthermore find that theoretical consistency of the model enforces
\begin{equation}\label{eq:Apconditiononcouplings}
    g_p\sim g_n\sim g_\chi,
\end{equation}
which constitutes a much stronger bound on $g_\chi$ than either unitarity or dark matter self-interactions, due to the bounds on $g_{p/n}$. For this reason, in \cref{sec:modelAp}, we will not discuss the SIDM bounds on the model.
In \cref{app:doubleaxialvectormodel}, we explain why obtaining $g_\chi \gg g_{n},g_p$ is very challenging, even in elaborate UV completions. 
With the condition in \eqref{eq:Apconditiononcouplings}, we find that direct detection signals are extremely suppressed  for $m_{A'} \lesssim m_B$ (here $m_B$ is the B-meson mass) due to a combination of bounds from SN1987A and rare meson decays on $g_{p/n}$.  For $m_{A'} \gtrsim m_B$, the most relevant bound comes from LHC searches for the new colored fermions needed for anomaly cancellation. 

In order to apply the above bounds to the direct detection rates,  we next derive the relevant non-relativistic Hamiltonian  
for the DM-proton interactions in \eqref{eq:UV_axion}, \eqref{eq:UV_axion2} and \eqref{eq:UV_vector}. In momentum space, they are:
\begin{align}
    \mathcal{H}_\phi &= - \frac{g_\chi g_{p}}{m_p} \frac{\mathbf{q}\cdot \mathbf{S}_{p}}{q^2+m_\phi^2}  e^{i\mathbf{q}\cdot \mathbf{r}} \label{eq:def_phi} \\
    \mathcal{H}_a &= - \frac{g_\chi g_{p}}{m_p m_\chi } \frac{(\mathbf{q}\cdot \mathbf{J_{\chi}})(\mathbf{q}\cdot \mathbf{S}_{p})}{q^2+m_a^2}   e^{i\mathbf{q}\cdot \mathbf{r}} \label{eq:def_a} \\
    \mathcal{H}^{\rm light}_{A'} &= \frac{4 g_\chi g_{p}}{m^2_{A^\prime}} \frac{(\mathbf{q}\cdot\mathbf{J}_\chi) (\mathbf{q}\cdot\mathbf{S}_p)}{{q^2}}  e^{i\mathbf{q}\cdot \mathbf{r}}   
    \label{eq:def_Ap_light}\\
    \mathcal{H}^{\rm heavy}_{A'} &= - 4 g_\chi g_p   \frac{\mathbf{J_{\chi}}\cdot \mathbf{S}_{p}}{{m_{A'}^2}}  e^{i\mathbf{q}\cdot \mathbf{r}},    \label{eq:def_Ap_heavy}
\end{align}
where $\mathbf{S}_{p}$ and $\mathbf{J}_{\chi}$ are respectively the proton and DM spin operators. $\mathbf{q}$ is the momentum exchanged, with $q\equiv |\bfq|$.  Here we have separated the $A'$ mediator into two cases: \eqref{eq:def_Ap_light} corresponds to the light mediator limit where $m_{A'}^2 \ll q^2$, while \eqref{eq:def_Ap_heavy} is the dominant operator in the heavy mediator limit where $m_{A'}^2 \gg q^2$. 
Due to the Goldstone boson equivalence theorem, $\mathcal{H}^{\rm light}_{A'}$ and $\mathcal{H}_a$ are identical, up to an overall redefinition of the coupling constants. 
Furthermore, the light mediator limit $m_{A'}^2 \ll q^2$ predicts extremely small direct detection cross sections if $g_{p/n} \sim g_\chi$.
In the bulk of this paper, we therefore restrict our discussion to \eqref{eq:def_phi}, \eqref{eq:def_a} and \eqref{eq:def_Ap_heavy}, and address \eqref{eq:def_Ap_light} in \cref{app:doubleaxialvectormodel} for completeness.
The Hamiltonians for the DM-neutron couplings have analogous forms.

\begin{table}[]
    \centering
    \begin{tabularx}{0.48\textwidth}{c|XX}
         & \multicolumn{1}{c}{$g_{p,n}$} & \multicolumn{1}{c}{$g_\chi$} \\\hline
         $\phi$ & Meson/SN (sec.~\ref{sec:alpSMbounds})& SIDM (sec.~\ref{sec:SIDMConstraints})\\
         $a$ & Meson/SN (sec.~\ref{sec:alpSMbounds})& SIDM/unitarity (sec.~\ref{sec:SIDMConstraints})\\
         $A'$& LHC (sec.~\ref{sec:modelAp})& gauge invar. (app.~\ref{app:AxialVectorModel} and \ref{app:doubleaxialvectormodel})\\
    \end{tabularx}
    \caption{Overview of most relevant constraints on $g_{p,n}$ and $g_\chi$ for each benchmark model, along with the sections where they are discussed.}
    \label{tab:boundssummary}
\end{table}

The scattering calculations are similar for these three models, and we therefore find it convenient to introduce the following shorthand notation:
\begin{align}
\mathcal{H} &= - \frac{g_\chi g_{p}}{q_0^2+m_{\rm med}^2} 
 F_{\rm med}(\mathbf{q}) \, \mathcal{O}\left(\mathbf{J}_\chi, \mathbf{q}\right)\cdot \mathbf{S}_{p} \, e^{i\mathbf{q}\cdot \mathbf{r}},\label{eq:def_general}
\end{align}
with $m_{\rm med}$ the mass of the mediator ({\emph i.e.},~$m_\phi$, $m_a$ or $m_{A'}$). The form factor is defined as 
\begin{equation}
    F_{\rm med}(\mathbf{q}) = \frac{q_0^2 + m_{\rm med}^2}{q^2 + m_{\rm med}^2}. \\
\end{equation}
The reference momentum $q_0$ is a convention, and is usually chosen to represent the typical momentum being exchanged in the collision. We choose $q_0 =m_\chi v_0$ with $v_0=220$ km/s being the typical dark matter velocity in the Earth's local neighborhood. 
The operator $\mathcal{O}$ for each model is given by
\begin{align} 
\mathcal{O}_\phi\left(\mathbf{J}_\chi, \mathbf{q}\right) &= \frac{\bfq}{m_p} \label{eq:Ophi} \\
 \mathcal{O}_a\left(\mathbf{J}_\chi, \mathbf{q}\right) &= (\mathbf{J}_\chi \cdot \bfq) \frac{\bfq}{m_p m_\chi} \label{eq:Oa} \\
\mathcal{O}_{A'}^{\rm heavy}\left(\mathbf{J}_\chi, \mathbf{q}\right) &= 4 \mathbf{J}_\chi.
\label{eq:OAprime} 
\end{align}

It is conventional to represent the sensitivity of direct detection experiments in terms of the DM-proton cross section, as if the proton were a free particle. In the $m_{\text{med}}\to 0$ limit, however, this quantity is divergent and we instead define the following reference cross sections to parametrize the reach:
\begin{align}
   \bar{\sigma}_\phi &= \frac{g_p^2 g_\chi^2 } {2\pi m_p^2} \frac{v_0^2\mu_{p\chi}^4  }{ (q_0^2+m_\phi^2)^2}\label{eq:refcrosssecphi} \\
    \bar{\sigma}_a &= \frac{g_p^2 g_\chi^2} {3\pi m_p^2 m_\chi^2} \frac{ v_0^4 \mu_{p\chi}^6}{(q_0^2+m_a^2)^2} \label{eq:refcrossseca} \\
      \bar{\sigma}_{A'}^{\rm heavy} &= 3\frac{g_p^2g_\chi^2}{\pi}\frac{\mu_{p\chi}^2}{m_{A'}^4}, \label{eq:refcrosssecAp} 
\end{align}
where $\mu_{p\chi}$ is the DM-proton reduced mass 
\begin{align}
\mu_{p\chi}\equiv \frac{m_\chi m_p}{m_\chi+m_p}.
\end{align}
These definitions reproduce the DM-proton cross section in the limit where the mediator mass $m_{\rm med} \gg q_0$. The operator in \eqref{eq:OAprime} and cross section in \eqref{eq:refcrosssecAp} are the most analogous to those that are commonly used for spin-dependent scattering of WIMPs.
The $q$ dependence of the operators in \eqref{eq:Ophi} and \eqref{eq:Oa} on the other hand leads to the different powers of $\mu_{p \chi}$ in the reference cross sections in \eqref{eq:refcrosssecphi} and \eqref{eq:refcrossseca}, which will impact the figures in \cref{sec:discussion}.



In our discussion, we will find it convenient to choose two benchmarks:
\begin{align}
    m_{a,\phi} &=0.3 \times q_0,\; \text{(light mediator benchmark)}
\label{eq:benchmarkvalueslight}\\
    m_{a,\phi} &=3 \times q_0,\; \text{(heavy mediator benchmark)}\label{eq:benchmarkvaluesheavy}
\end{align}
for the spin 0 mediators ($\phi,a$).
They approximately represent the optimal choices, from the point of view of maximizing the available parameter space for direct detection.
This means that we always consider $m_{a,\phi}\ll m_\chi$, even in the ``heavy'' mediator regime.
As such, the $\phi,a$ cannot decay to the dark matter, which will be relevant when we consider the bounds on these particles.

For the $A'$, we will focus on the benchmark $m_{A'} = 10$ GeV. Here the ${A'}$ is too heavy to be produced in exotic $B$ meson decays and the most stringent bound is provided by LHC searches for new colored fermions that are required to cancel the anomalies of the $U(1)'$. 
This always lands us in the heavy mediator regime for scattering, $m_{A'} \gg q_0$. As explained above, we do not consider the light mediator limit for the $A'$ model, except briefly in \cref{app:doubleaxialvectormodel}.





\section{Bounds on spin-0 mediators\label{sec:scalarbounds}}

\subsection{Mediator -- SM coupling \label{sec:alpSMbounds}}

Any field with substantial couplings to protons and/or neutrons must have inherited it from a coupling to SM quarks and/or gluons.
Such couplings in turn induce exotic meson decay constraints through electroweak penguin diagrams, which are tightly constrained by experiments.
For concreteness, we consider the operator
\begin{equation}\label{eq:uvdefinition}
 \mathcal{L} \supset  c_{GG}\frac{\alpha_s}{4\pi} \frac{a}{f_a} G^{a}_{\mu\nu} \tilde{G}^{a,\mu\nu},   
\end{equation}
with $\tilde G^{a,\mu\nu}\equiv\frac{1}{2}\epsilon^{\alpha\beta\mu\nu}G^a_{\alpha\beta}$, and analogously for $\phi$. We take $f_a = 1$ TeV and $\Lambda=4\pi f_a$ to be the scale for the UV boundary conditions of this theory. 
At this scale, the couplings of $a/\phi$ to all other SM particles apart from the gluons are assumed to vanish. 

\begin{figure*}[tb]
\centering
\includegraphics[width=0.48\linewidth]{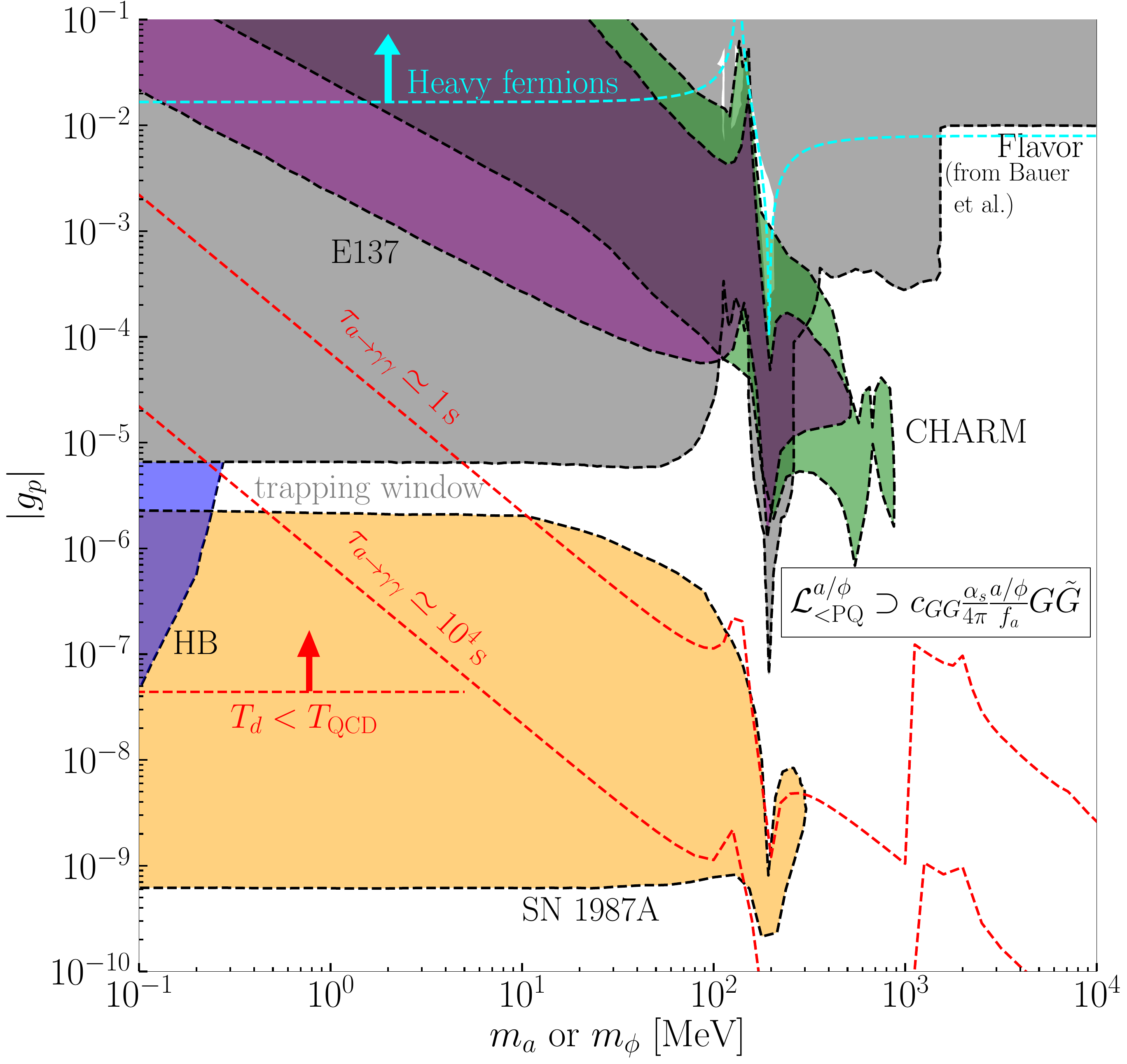}\hfill ~~~~
\includegraphics[width=0.48\linewidth]{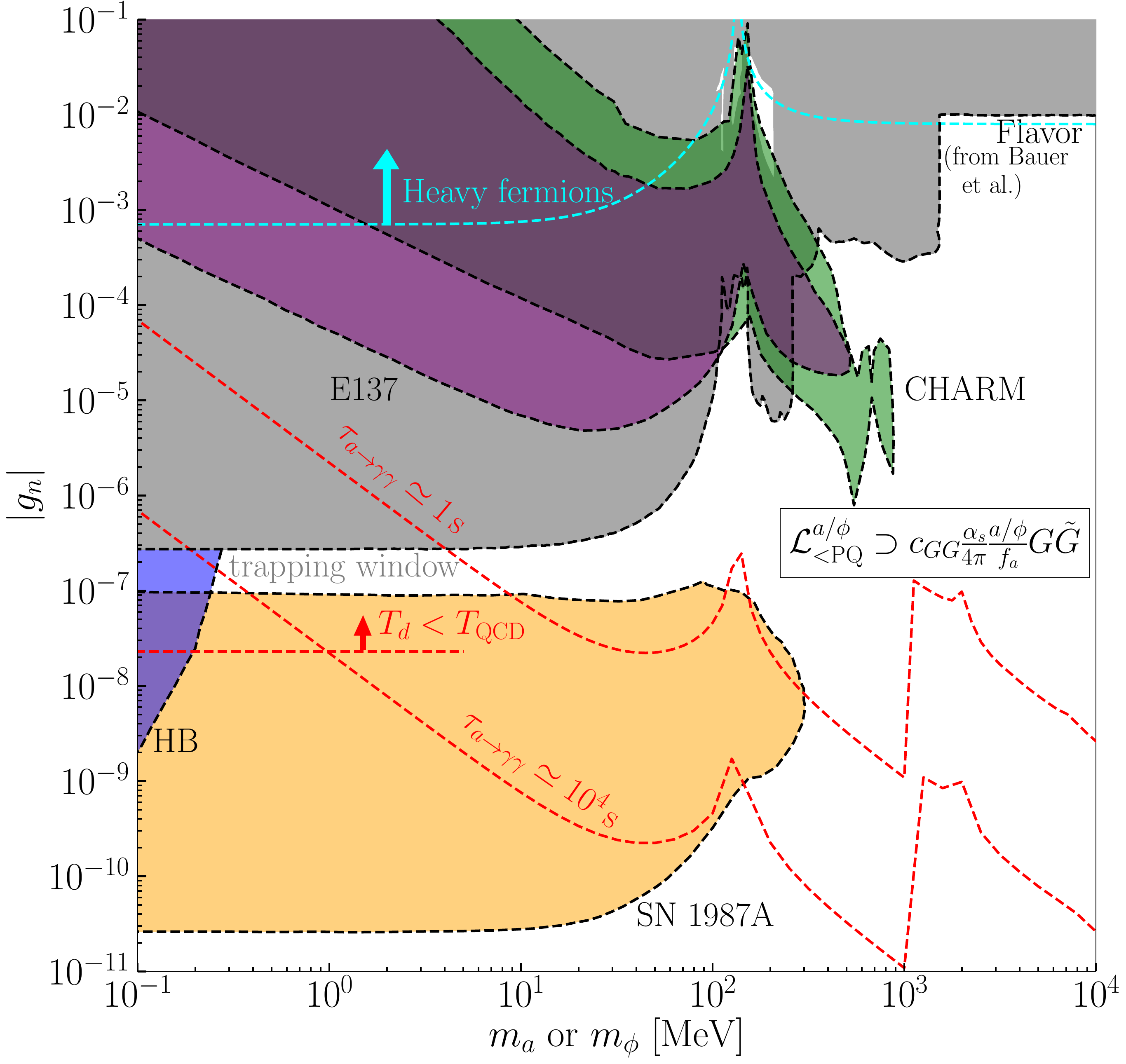}
\caption{The left (right) panel shows constraints on the axion-nucleon coupling $g_p$ ($g_n$) in the plane of the mediator $m_a$ or $m_\phi$ arising from \eqref{eq:uvdefinition}. The main bounds are from flavor observables in rare meson decay experiments~\cite{Bauer:2020jbp} (grey), the beam-dump experiment CHARM~\cite{CHARM:1985anb, Jerhot:2022chi} (green), the electron beam-dump experiment E137~\cite{Dolan:2017osp} (purple), horizontal branch (HB) stars~\cite{Carenza:2020zil} (blue) and supernova SN1987A~\cite{Lella:2022uwi, Lella:2023bfb} (yellow). The horizontal dashed red line shows the value of the axion-nucleon coupling required for the spin-0 mediator to fall out of equilibrium with the pions at a decoupling temperature of 70 MeV~\cite{DiLuzio:2021vjd}. Above this line and for a massless mediator, the mediator contributes $\Delta N_{\rm eff} > 0.36$ after $e^+e^-$ annihilation. The sloping dashed red lines show the lifetime needed ($\tau_{a\gamma\gamma}\lesssim 10^4$ s) to evade BBN constraints on late-time energy injection. The region enclosed between the 1s and $10^4$s lines indicate the region where the $\phi/a$ could contribute to $N_{\rm{eff}}$ during BBN and recombination. The dashed cyan line denotes the bound in ~\eqref{eq:heavy_fermions} from the presence of heavy colored fermions, coming from UV completing the effective interaction  \eqref{eq:uvdefinition} in a KSVZ model (see \eqref{eq:KSVZ}). } 
\label{fig:Constraints_med}
\end{figure*}

We use the formalism in \cite{Bauer:2021mvw,Bauer:2020jbp,GrillidiCortona:2015jxo} to map the gluon coupling in \eqref{eq:uvdefinition} to the couplings with protons and neutrons: 
\begin{align}
g_{p}\approx& \frac{m_p}{f_a} c_{GG} \left(g_0+g_A \frac{m_\pi^2}{m_\pi^2-m_a^2}\frac{m_d-m_u}{m_u+m_d}\right)\label{eq:gp_alp}\\
\approx& ~8.11\times 10^{-4}\times c_{GG},\label{eq:gp_alp_limit}\\
g_{n}\approx& \frac{m_p}{f_a} c_{GG} \left(g_0-g_A \frac{m_\pi^2}{m_\pi^2-m_a^2}\frac{m_d-m_u}{m_u+m_d}\right)\nonumber\\
&+c_{GG} \delta_{RG}(f_a)\label{eq:gn_alp}\\
\approx& -3.50\times 10^{-5} \times c_{GG} ,\label{eq:gn_alp_limit}
\end{align}
where \eqref{eq:gp_alp_limit} and \eqref{eq:gn_alp_limit} are valid in the limit where \mbox{$m_a\ll m_\pi$}.
The phenomenological couplings $g_A= 1.2754$ and $g_0= 0.440$ are respectively derived from $\beta$-decay measurements \cite{ParticleDataGroup:2024cfk} and lattice QCD results \cite{Liang:2018pis} (see \cite{Bauer:2021mvw} for details).
The parameter $\delta_{RG} (f_a) \approx-1.56\times 10^{-5}$ contains the corrections from the renormalization group running from $\Lambda$ down to the IR scale $\mu_0=2$ GeV \cite{Bauer:2021mvw}. 
For $g_p$, these corrections are always subleading to the other terms and can be safely neglected.
For $g_n$, there is an accidental cancellation between the first two terms in \eqref{eq:gn_alp}, such that the renormalization group running has an $\mathcal{O}(1)$ impact.
As we will see in \cref{sec:scatteringcalculation}, most materials used in low threshold direct detection experiments are primarily sensitive to $g_p$ rather than $g_n$.\footnote{
 For direct detection targets sensitive primarily to $g_n$, the cancellation in \eqref{eq:gn_alp_limit} will imply reduced sensitivity. 
By considering the axion couplings to quarks, $\partial^\mu a \bar q \gamma_\mu \gamma_5 q$, it is likely that a model could be constructed for which the bounds on $g_n$ from meson decays are somewhat relaxed. In such a model, the upper bounds on the direct detection rate would be slightly weaker than what one would obtain from the right-hand panel of \cref{fig:Constraints_med}. However, this only applies for materials with odd number of neutrons, and we do not expect our conclusions to change qualitatively.
}\\

At energy scales lower than $f_a$, the $a$-photon (or $\phi$-photon) coupling is generated
\begin{equation}
    \mathcal{L}\supset c_{\gamma\gamma}\frac{\alpha}{4\pi}\frac{a}{f_a}F_{\mu\nu}\tilde{F}^{\mu\nu},
\end{equation}
where $\alpha$ is the fine-structure constant. 
Below the QCD scale, the $a$-photon coupling obtains sizeable contributions from QCD non-perturbative effects~\cite{Bauer:2021mvw}
\begin{equation}\label{eq:Cgammagamma}
    c_{\gamma\gamma} \simeq -1.92 c_{GG} - c_{GG}\frac{m_a^2}{m_\pi^2-m_a^2}\frac{m_d - m_u}{m_d+m_u}.
\end{equation}
The above expression neglects the subleading contributions from operators of the form $\partial^\mu a \bar q\gamma_\mu \gamma_5 q$, which are generated in the RGE evolution from the UV scale ($\Lambda$) to the weak scale.\footnote{Meanwhile, for $m_a \gg \Lambda_{\rm QCD}$, $c_{\gamma\gamma}$ is suppressed as it only receives the contributions from the RG-running at 1-loop;  we take the cut-off threshold between the two regimes to be at $m_a = 1$ GeV. This explains the kink in the red line for $\tau_{a\to\gamma\gamma}\simeq 1$s in \cref{fig:Constraints_med}.} The $a$-photon coupling will be important for some bounds from astrophysics, cosmology, and beam dumps, which we discuss below.

The constraints on $c_{GG}$, including the impact of radiative corrections, were analyzed in detail in \cite{Bauer:2021mvw,Bauer:2020jbp}.
We use \eqref{eq:gp_alp} and \eqref{eq:gn_alp} to map their bounds from flavor observables onto $g_p$ and $g_n$. This is shown in \cref{fig:Constraints_med}. In the mass range \mbox{$0.1~\text{GeV} \lesssim m_a \lesssim 1~\text{GeV}$}, we further included a reinterpretation \cite{Jerhot:2022chi} of a search performed by the CHARM beam-dump experiment \cite{CHARM:1985anb}. 
This latter bound is subject to a number of caveats, which we comment on in \cref{app:charm}. In the same vein, slightly weaker limits are afforded by the beam-dump experiment E137 which fired a $20$ GeV electron beam onto aluminum plates immersed in cooling water, see purple region in the figure. For this bound, we rescaled the results presented in \cite{Dolan:2017osp} using \eqref{eq:gp_alp} and \eqref{eq:Cgammagamma}.  

We also include cooling bounds from Supernova SN1987A, as the emission of $a/\phi$ particles could lead to a modified neutrino spectrum observed in the Kamiokande II detector \cite{Kamiokande-II:1987idp}. This leads to a bound (shown in yellow in \cref{fig:Constraints_med}) on the $a/\phi$-nucleon coupling, which we obtain from~\cite{Lella:2022uwi,Lella:2023bfb}. 
The bound relies on the emitted luminosity into the $a/\phi$ particles, which depends on a quadratic combination of $g_{p}$ and $g_n$, given as a fitting formula in Eq.~(11) of \cite{Lella:2022uwi}. We apply this for the couplings in \eqref{eq:gp_alp} and \eqref{eq:gn_alp}, interpolating over the coefficients of the fitting formula.  
In our model, for $m_a \ll m_\pi$, the ratio $g_p/g_n \gg 1$ so that the bound appears stronger on $g_n$ than on $g_p$. For $m_a > m_\pi$, then $g_p/g_n \lesssim 1$ causing the bound on $g_p$ to become stronger than that on $g_n$. We emphasize that the SN1987A bound has $O(1)$ uncertainties, as discussed in detail in~\cite{Lella:2023bfb}.

For masses $m_a\lesssim 0.5$ MeV, we show the cooling bound from horizontal branch stars, whereby the cooling mechanism induced by the $a$-photon coupling would reduce the ratio of horizontal branch stars to red giant branch stars in globular clusters~\cite{Carenza:2020zil}. In our model, the $a$-photon coupling is proportional to $c_{GG}$ and thus maps directly to constraints on the $a$-nucleon couplings $g_p$ and $g_n$.


For $100\;\mathrm{keV}\lesssim m_a \lesssim 100$ MeV, there currently is a narrow range of allowed couplings $2 \times 10^{-6} \lesssim g_p \lesssim 7 \times 10^{-6} $, in between the meson and supernova bounds in \cref{fig:Constraints_med}, and similarly for the neutron coupling.
Here the mediator cannot free stream out of the proto-neutron star, which is why the SN1987A cooling bound cuts off. We will refer to this region as the ``trapping window''.
This window will prove to be important because it allows for larger direct detection cross sections, but we caution that this parameter space may be subject to additional bounds from both cosmology and astrophysics, which require more detailed study. For instance, it is possible that future work on diffusive energy transport by axion-like particles would strengthen the supernova bounds to the extent that the trapping window will be closed. (See \cite{Fiorillo:2025yzf} for such an analysis for axion-like particles which only couple to photons.) 

Next, we discuss the cosmology. In the trapping window, the coupling is large enough that the mediator is still thermalized with the SM bath after the QCD phase transition, until a decoupling temperature of $T_{\rm d} \simeq 20$ MeV. This implies that the mediator would contribute to the effective number of neutrinos ($\Delta N_{\rm eff}$). However, the mediator also decays to two photons $a\to\gamma\gamma$ through the photon coupling. As long as the decay time is less than $10^4$ s, it is not subject to constraints on late-decaying particles from photodissociation of light elements~\cite{Forestell:2018txr}. 
However, even if the decay is in the range 1~s to $10^4$ s, it still occurs during BBN, thus dumping energy back into the photons. This will reduce $\Delta N_{\rm eff}$, eventually reaching a constant negative value once the decay comes fully into thermal equilibrium.  
Thus, the mediator could still modify the expansion rate during BBN, impacting BBN abundances as well as contributing to $\Delta N_{\rm eff}$ during recombination. 
Since the mediator freeze-out and subsequent decay leads to a time-dependent $\Delta N_{\rm eff}$, a dedicated analysis of this model is necessary to evaluate the constraints, for example similar to \cite{Berlin:2019pbq}. 

Here we perform a simple estimate to determine where the BBN bounds may be relevant.
Assuming an instantaneous energy dump into the photons once $a \to \gamma\gamma$ is thermalized, we estimate that $m_a\gtrsim 10$ MeV is needed to avoid tension at $2\sigma$ from a combined analysis of BBN and CMB measurements giving $|\Delta N_{\rm eff}| \lesssim 0.3$~\cite{Yeh:2022heq}. This is consistent with the results of \cite{Depta:2020wmr} which have performed dedicated calculations on the impact of axion-like particles on the light element abundances and $N_{\rm eff}$ at CMB to find $m_a \gtrsim 6 $ MeV. However, \cite{Depta:2020wmr} only considers the axion-photon coupling and is therefore not directly applicable to our benchmark model. Furthermore, the bounds can be weakened by reheating at a lower temperature, allowing for extra contributions to $N_{\rm eff}$ or allowing a non-vanishing chemical potential of neutrinos.  We therefore do not impose any bounds in the trapping window from cosmology. Instead in \cref{fig:Constraints_med}, we show in dashed red lines corresponding to $\tau_{a\to\gamma\gamma} \simeq 10^4$ s and $\tau_{a\to\gamma\gamma} \simeq 1 $ s to indicate approximately where the above considerations about the cosmological history might play a role.

Finally, we remark that a consistent UV completion for \eqref{eq:uvdefinition} is afforded by the KSVZ model
\begin{equation}
    \mathcal{L}_{\rm KSVZ} \supset y_\star \Phi \bar{\psi}_L\psi_R + \text{h.c.}
\end{equation}
where $\psi$ is a set of $N_\psi$ heavy, colored, vector-like fermions in the fundamental representation of $SU(3)_c$ and $y_\star$ is their Yukawa coupling to the Peccei-Quinn field $\Phi$. $\Phi$ spontaneously acquires the vev $\Phi_0 = v_a e^{i a/v_a}/\sqrt{2}$ and after performing a chiral rotation $\psi \to e^{-i\gamma^5/(2v_a)}\psi$ with $v_a = f_a/(2 c_{GG})$, a mass term is generated for the fermions
\begin{equation}
    m_\psi = \frac{y_\star N_\psi f_a}{2\sqrt{2} c_{GG}},
    \label{eq:KSVZ}
\end{equation}
and the anomaly term is generated as in \eqref{eq:uvdefinition}.
Requiring that the colored fermions have mass $m_\psi \gtrsim 2$ TeV (for a review of the LHC bounds on vector like fermions, see \cite{CMS:2024bni}), $y_\star < 4\pi$ and $N_\psi \leq 10$, this implies a bound of $c_{GG}\lesssim 20$ or equivalently
\begin{equation}
\label{eq:heavy_fermions}
|g_p| \lesssim 2\times 10^{-2} \quad \text{and}\quad |g_n|\lesssim 7\times 10^{-4}
\end{equation}
for $m_a\ll m_\pi$. This bound is indicated by the the cyan line on \cref{fig:Constraints_med}. 
In principle, these bounds can be evaded by taking larger $N_\psi$, but only at the expense of introducing a Landau pole in the strong coupling before the scale of grand unification.

\subsection{Mediator -- DM coupling}

\label{sec:SIDMConstraints}

Next we consider constraints on the mediator coupling to DM, $g_\chi$, arising from self-interacting dark matter bounds. 
Frequent dark matter self-interactions will drive the velocity distribution towards an isothermal one, leading to more spherical halos with flatter central densities, as compared to collisionless dark matter \cite{Spergel:1999mh, Kaplinghat:2015aga}. 
Various astrophysical systems (dwarf galaxies, low surface brightness galaxies, galaxy clusters, etc.) spanning a range of scales provide different constraints on the strength of the dark matter self-interactions (see~\cite{Tulin:2017ara} for a review).
We will use the constraint derived from galaxy groups \cite{Sagunski:2020spe}
\begin{equation}
\label{eq:SIDM_bound}
    \sigma/m_\chi \lesssim 1.1 \, \text{cm}^2 / \text{g},
\end{equation}
for an average relative velocity of $v\simeq 0.005 c$ (1430$\text{ km s}^{-1}$).

To apply the bound in \eqref{eq:SIDM_bound} for models with a light mediator, one must down-weight scattering events with low momentum transfer, as they do not significantly alter the DM velocity profile.
We do so by identifying the cross section in \eqref{eq:SIDM_bound} with the viscosity cross section
\begin{equation}
\sigma_V \equiv \int d\Omega\, \frac{d\sigma}{d\Omega}  (1-\cos^2\theta).
\end{equation}
This definition is suitable for indistinguishable particles, as it regulates divergences in both the forward and backward scattering for low mass mediators.

Below we will report the viscosity cross-sections for each of the spin-0 mediators, and derive a constraint on $g_\chi$ by comparing with \eqref{eq:SIDM_bound}.
We work in the non-relativistic limit $v \ll 1$ and present the results in terms of the quantity 
\begin{equation}
    R \equiv \frac{m_\chi v} { m_{\rm med}},
\end{equation}
where $v$ is the relative velocity of the colliding DM particles.

\begin{figure*}[tb]
\centering
\includegraphics[width=0.49\linewidth]{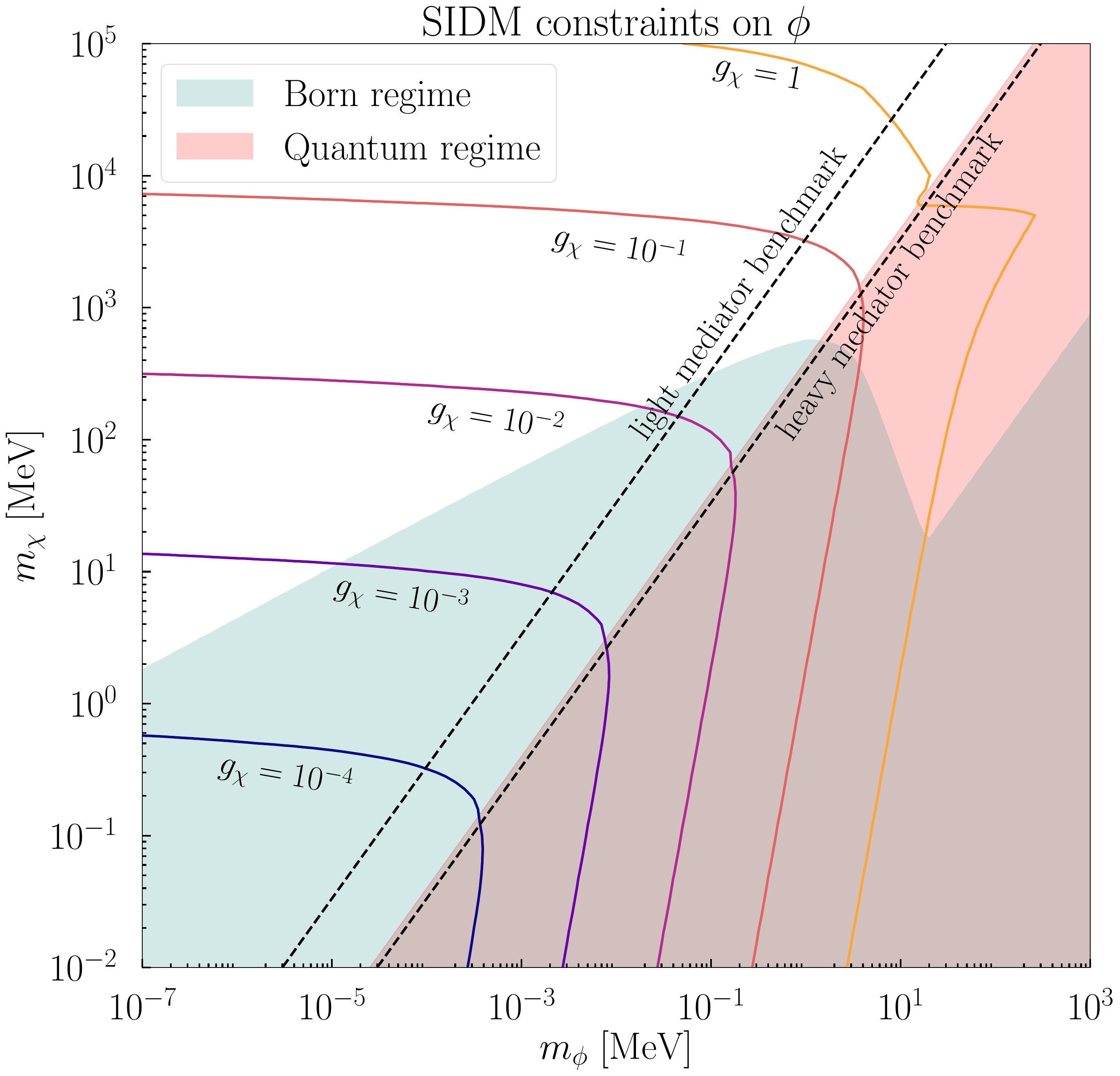} 
\includegraphics[width=0.49\linewidth]{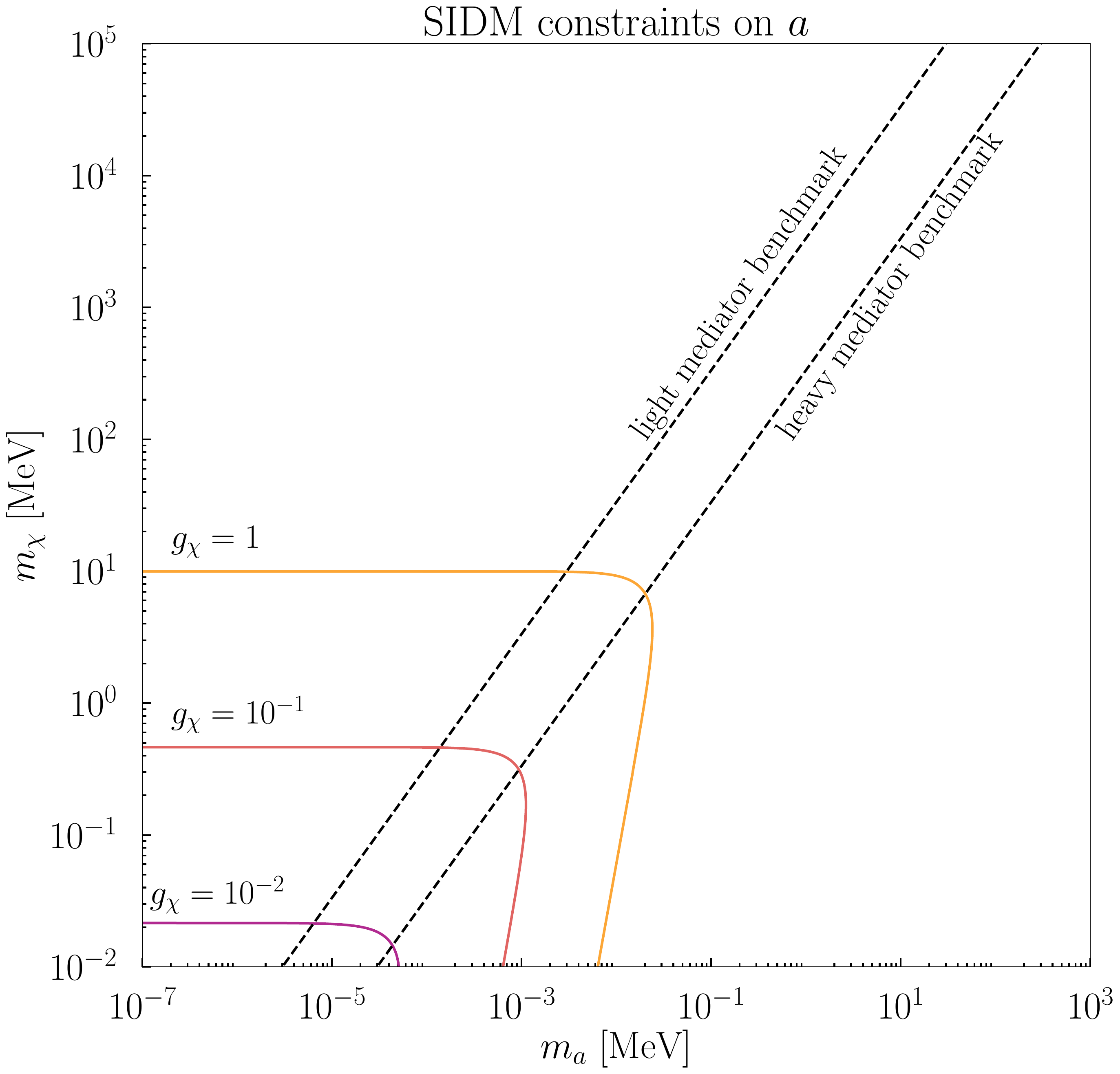}
\caption{Contours for the self-interacting DM bound on $g_\chi$. The dashed lines correspond to the light ($m_{\rm red} = 0.3 m_\chi v_0$) and heavy ($m_{\rm red} = 3 m_\chi v_0$) mediator direct detection benchmarks of \eqref{eq:benchmarkvalueslight} and \eqref{eq:benchmarkvaluesheavy} ({\bf left}) Bounds for the $\phi$ mediator. The red region corresponds to the quantum regime and the teal region to the regime where the Born approximation is valid.  ({\bf right}) Bound for the $a$ mediator.  
}
\label{fig:Constraints_SIDM}
\end{figure*}

\vspace{-3mm}

\subsubsection{$\phi$ mediator\label{sec:SIDMphi}}

From the perspective  of the dark matter self-interactions, the $\phi$ model is identical to the light dark matter model often considered for spin-independent scattering~\cite{Knapen:2017xzo,Tulin:2017ara}.
At tree-level, the viscosity cross section is
\begin{multline}
    \label{eq:SIDM_phi}
    \sigma_V^\phi  \simeq  \frac{g_\chi^4}{2\pi m_\chi^2 R^2 (R^2+2)v^4} \Bigg [ -5 R^2 (R^2+2) \, +   \\  2(R^4+5R^2+5) \log(1+R^2)  \Bigg ] \stackrel{R \ll 1}{\approx} \frac{ g_\chi^4 m_\chi^2}{12 \pi m_\phi^4}.
\end{multline}
Above we have also given the behavior for $R \ll 1$. 

The expression in \eqref{eq:SIDM_phi} is only valid in the Born approximation where $(g^2_\chi m_\chi)/(4\pi m_\phi) \ll 1$. When the scalar mediator $\phi$ is sufficiently light that $(g^2_\chi m_\chi)/(4\pi m_\phi) \gg 1$, the dark matter particles interact via multiple exchanges of the mediator resulting in non-perturbative effects that need to be resummed \cite{Sommerfeld:1931qaf}. 
In this case, to obtain a reliable computation of the bound, we use the results of \cite{Colquhoun:2020adl}, which gives analytic expressions in all the relevant regimes for scattering. 
The regime $R \gtrsim 1$ corresponds closely to the (semi)-classical regime, defined as $m_\chi v/ (2 m_\phi) \gg 1$, and here we use Eq.~(86) in \cite{Colquhoun:2020adl}. The regime $ R \lesssim 1$ corresponds to the quantum regime,  
where $S$-wave scattering dominates. The result can be well-approximated by solving for scattering in a Hulth\'en potential~\cite{Tulin:2013teo,Colquhoun:2020adl}, which gives:
\begin{equation}
\label{eq:SIDM_phi_quantum}
    \sigma_V^\phi \simeq \frac{16\pi}{3 m_\chi^2 v^2}\sin^2 \delta_0,
\end{equation}
where $\delta_0$ is the phase shift for the $s$ partial wave that is given in terms of Gamma functions by 
\begin{align}
\delta_0 &= \text{arg}\left ( \frac{i\Gamma(\lambda_+ + \lambda_- - 2)}{\Gamma(\lambda_+)\Gamma(\lambda_-)} \right )\\
\lambda_\pm &= 1 + i \frac{m_\chi v}{2 m_\phi n} \pm \frac{m_\chi v}{2 m_\phi n}\sqrt{ \frac{g_\chi^2 m_\phi}{\pi m_\chi v^2} n - 1},
\end{align}
and $n = \sqrt{2 \zeta(3)}$ is a constant fixed by matching to the  $R\ll 1$ limit of \eqref{eq:SIDM_phi}. 

Using the above results, we obtain upper bounds on $g_\chi$ as a function of $(m_\chi, m_\phi)$ by saturating the bound given in \eqref{eq:SIDM_bound}. These are given as the different contours in the left panel of \cref{fig:Constraints_SIDM}. 
The teal shaded region indicates the regime where the Born approximation (and hence \eqref{eq:SIDM_phi}) is valid. 
Note that the change in slope of the teal region around $m_\phi \approx$ 10 MeV arises because we require $g_\chi < 1$. 
The red shaded region indicates the quantum regime, where we have used the Hulth\'en potential cross section. (The cross sections agree in the area where both regimes intersect.) 
In the quantum regime,  when the coupling is sufficiently strong that $g_\chi^2 m_\phi / (\pi m_\chi v^2) \gg 1$, there are multiple resonances and anti-resonances appearing in the parameter space due to the sinusoidal nature of \eqref{eq:SIDM_phi_quantum}. 
This can give multiple ranges of $g_\chi$ satisfying \eqref{eq:SIDM_bound}. 
The effect is most relevant in the upper right corner of the parameter space. In those cases, to be conservative, we always take the value of the first $g_\chi$ crossing where \eqref{eq:SIDM_phi_quantum} starts to violate \eqref{eq:SIDM_bound} as our bound, and neglect any subsequent resonant dips. 
In practice, we expect narrow ranges of $g_\chi$ corresponding to resonances to be excluded if we had considered the full spectrum of SIDM bounds, obtained from a suite of astrophysical systems spanning a spectrum of velocities and densities.
To guide the reader's eye, the left panel of \cref{fig:Constraints_SIDM} also shows the two direct detection benchmarks (light and heavy mediator) we introduced in \eqref{eq:benchmarkvalueslight} and \eqref{eq:benchmarkvaluesheavy}.
\vspace{-2mm}
\subsubsection{$a$ mediator}
Repeating the same procedure, we calculated the viscosity cross section with the $a$ mediator at tree-level in quantum field theory and took the non-relativistic limit
\begin{multline}
    \label{eq:SIDM_a}
    \sigma_V^a \simeq \frac{g_\chi^4}{64\pi m_\chi^2 R^6 (R^2+2)}\Bigg [ 4(2R^2+3)^2 \log(1+R^2) \, + \\    
     (R^6 - 4 R^4 -30 R^2 -36)R^2\Bigg ] \stackrel{R \ll 1}{\approx} \frac{g_\chi^4 m_\chi^2 v^4}{240\pi m_a^4}.
\end{multline}
For the $a$ mediator, it was shown that terms in the non-relativistic potential that could potentially have given rise to Sommerfeld enhancement are suppressed by $m_\phi^2/m_\chi^2$ when the mediator is lighter than the DM particle~\cite{Agrawal:2020lea}, which is the case for both our light and heavy mediator benchmarks.  
The expression in \eqref{eq:SIDM_a} can therefore be applied throughout the parameter space we consider. The resulting bound on $g_\chi$ is shown in the right panel of \cref{fig:Constraints_SIDM}.

\subsubsection{SIDM summary}
\begin{figure}[t]
\includegraphics[width=\linewidth]{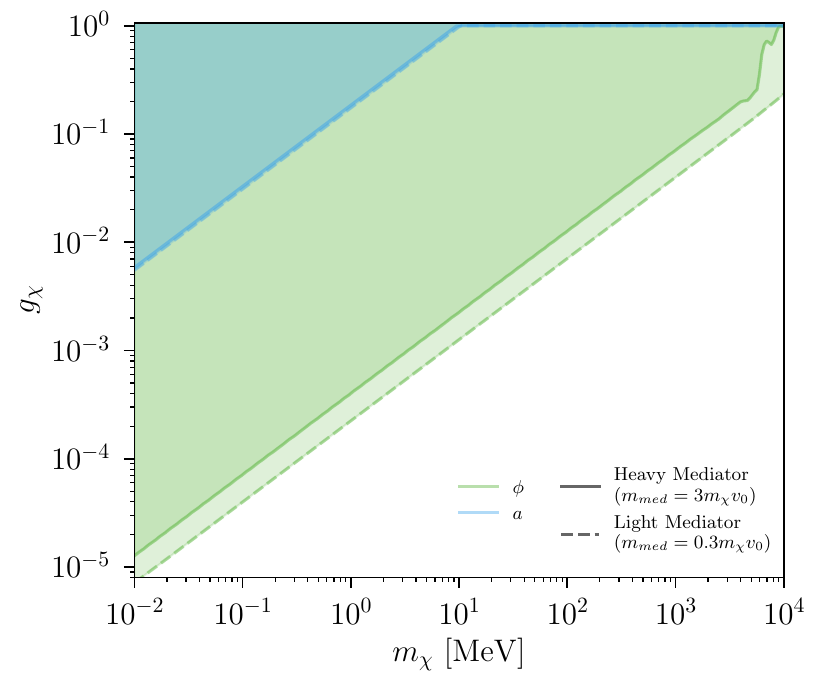}
\caption{SIDM constraints on $g_\chi$ for spin 0 mediators, assuming the light mediator ($m_{\rm med} = 0.3 m_\chi v_0$) and heavy mediator ($m_{\rm med} = 3 m_\chi v_0$) benchmarks.} 
\label{fig:Constraints_SIDM_summary}
\end{figure}

We summarize constraints on all mediators in \cref{fig:Constraints_SIDM_summary}, again for the two direct detection benchmarks in \eqref{eq:benchmarkvalueslight} and \eqref{eq:benchmarkvaluesheavy}. As $m_\chi$ grows, the SIDM bound becomes weaker (i.e.~$g_\chi$ increases) until it ceases to be a meaningful bound since we demand that the model remains perturbative. Here we require that $g_\chi \le 1$.
We conclude that SIDM constraints are particularly strong for the $\phi$ mediator, regardless of whether the mediator is in the ``heavy'' or ``light'' regime of direct detection.

It is important to bear in mind that these bounds are derived under the assumption that the self-interacting $\chi$ particle contributes 100\% of the dark matter density. 
If $\chi$ instead makes up only a small fraction of the dark matter ($\lesssim 5\%$), its effect on the dynamics of the dark matter halos is expected to be suppressed. (Although we note that for certain models, such as atomic dark matter, even an abundance of 5\% can have dramatic consequences~\cite{Gemmell:2023trd}).
Defining $f_{DM} = \Omega_\chi/\Omega_{\rm DM}$, we will consider both the case where this candidate is all of the DM ($f_{DM} =1$) and a sub-component ($f_{DM} = 0.05$) in \cref{sec:discussion}. In the sub-component scenario, we set  $g_\chi =1$, saturating our perturbativity constraint.

\section{Bounds on axial vector mediator\label{sec:modelAp}}

Any axial vector mediator model of the form in \eqref{eq:UV_vector} suffers from gauge anomalies at low energies, and therefore needs additional heavy particles which are chiral under the corresponding $U(1)^\prime$ symmetry. 
In general, such a self-consistent UV completion cannot to be taken for granted and will make important predictions for the infrared physics.

\begin{figure}
    \centering
    \includegraphics[width=\linewidth]{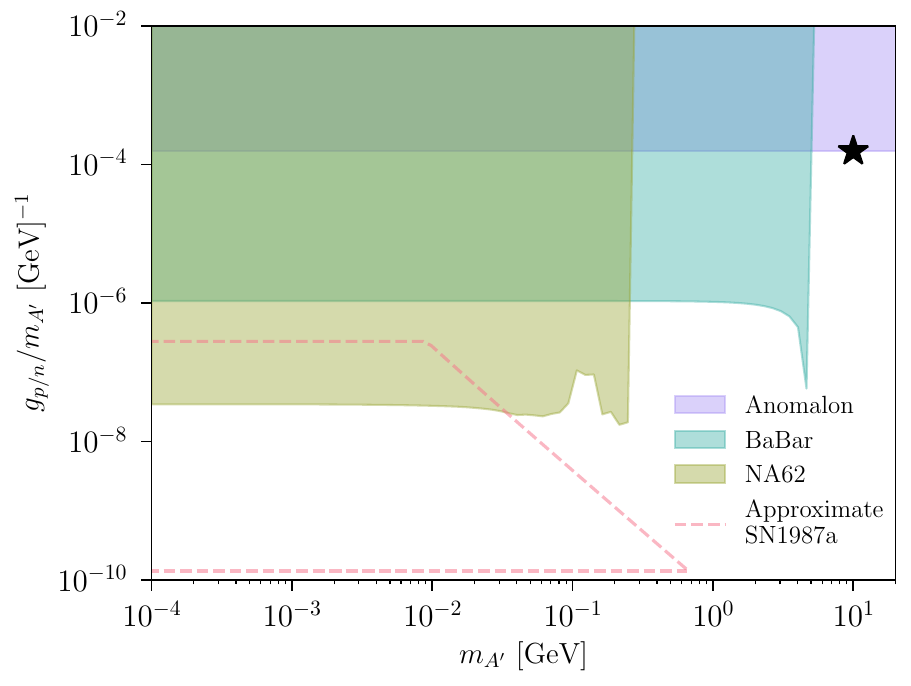}
    \caption{Bounds on $A'$ mediator from LHC limits on the anomalons (\eqref{eq:anomalonbound} and \cref{app:AxialVectorModel}), $B\to K \nu\bar \nu$ at BaBar, $K\to \pi \nu\bar \nu$ at NA62, and $A'$ emission during SN1987A (see text for details.) The star indicates the optimal benchmark value which we will adopt to compute the possible signals at direct detection.}
    \label{fig:Aprime_constraints}%
\end{figure}

Concretely, the minimal UV completion we construct in \cref{app:AxialVectorModel} has the couplings 
\begin{equation}\label{eq:anomalouslowE}
    \mathcal{L} \supset g' A'^\mu \left[ \sum_q \bar q \gamma_\mu \gamma_5 q - 2 \bar \nu \gamma_\mu P_R \nu - \frac{1}{2}  \bar \chi \gamma_\mu \gamma_5 \chi\right],
\end{equation}
with flavor-universal quark couplings and where $g'$ is the $U(1)'$ gauge coupling.
We assumed Dirac neutrinos and charged their right-handed component under the $U(1)'$, which gives rise to the second term. This interaction simplifies the meson decay bounds on the $A'$, as this makes $A' \to \bar{\nu} \nu$ the dominant decay mode at low masses. At higher masses, the $A'$ can also invisibly decay into two dark matter particles.\footnote{Without the neutrino coupling, the dominant decay will be into hadrons, if phase space is available.
In addition, the applicable NA62 and Belle II measurements would need to be recast. 
Such a study is outside the scope of this work.
While such a visible, hadronic decay mode would likely somewhat weaken the NA62 and Belle II bounds, our results for the dark matter scattering rate would remain unchanged, since the optimal parameter point for direct detection would remain unchanged in \cref{fig:Aprime_constraints}.}

To map the coupling to quarks onto the IR coupling to protons, we use the following relation for the hadronic matrix elements~\cite{Hill_2015}:
\begin{equation}
\begin{split}
    \langle & p(k')|\sum_{q = u,d,s} \bar q \gamma_\mu \gamma_5 q |p(k)\rangle \equiv  \\ & \bar{u}^{(p)}(k')\bigg[ F_{A'}^{(p)}(q^2)\gamma_\mu\gamma_5 + F_{P'}^{(p)}(q^2)\gamma_5 \frac{q_\mu}{2 m_p} \bigg]u^{(p)}(k),
\end{split} 
\end{equation}
where $F_{A'}^{(p)}$ and  $F_{P'}^{(p)}$ are form factors and an analogous equation holds for the coupling to neutrons $n$.
We only consider the first term since the second term is suppressed for sub-GeV DM, where $q \sim m_\chi v  \ll m_p$. In addition, we only need the $q \to 0$ behavior of the form factor for sub-GeV DM. 
We use the values extracted from \cite{Nocera_2014}, which gives
\begin{equation}\label{eq:formFactorAxialVec}
    F_{A'}^{(p)}(0) = F_{A'}^{(n)}(0) = 0.22(15).
\end{equation}
This leaves us with the prediction
\begin{equation}
\begin{split}
    g_{p,n} &\approx 0.22 \times g' \\
    g_\chi &=-\frac{1}{2} g'.
\end{split} \label{eq:thisisbad}
\end{equation}

As we will see below, bounds on $A'$ couplings to the SM impose $g_{p,n} \ll 1$, which makes the relationship in \eqref{eq:thisisbad} particularly challenging for the direct detection prospects. 
One may be tempted to assume the existence of a more general model which can accommodate a hierarchy of the form 
\begin{equation}\label{eq:thiswouldbebetter}
    g_{p,n} \ll g_\chi \sim 1,
\end{equation}
and a more sizable rate at direct detection. In \cref{app:doubleaxialvectormodel}, we show that such a generalization does exist that gives the light mediator limit $m_{A'}\ll q$ for scattering \eqref{eq:def_Ap_light}, but only at the expense of rather elaborate dark sector model building.
We are unable to construct a self-consistent model which can accommodate the hierarchy \eqref{eq:thiswouldbebetter} and where the heavy mediator ($m_{A'}\gg q$) scattering operator in \eqref{eq:def_Ap_heavy} is the dominant interaction.
Throughout the rest of the paper, we therefore assume that the relation \eqref{eq:thisisbad} holds, and refer the reader to \cref{app:doubleaxialvectormodel} for a discussion on more general models and their challenges. 
 
The interaction in \eqref{eq:anomalouslowE} implies a gauge anomaly in the IR theory, which means that the model must be UV completed with additional fermions - ``anomalons'' - at a scale low enough to render the theory unitary (see \cref{app:AxialVectorModel}).  
By demanding that these additional fermions are not in tension with LHC bounds, we arrive at the approximate bound 
\begin{equation}
g' \lesssim \frac{1}{\sqrt{2}}\frac{m_{A'}}{\mathrm{TeV}}. \label{eq:anomalonbound}
\end{equation}

The $A'$ moreover mixes with the hypercharge gauge field through
\begin{equation}\label{eq:kineticmixing}
    \mathcal{L}\supset \frac{1}{2}\epsilon F'^{\mu\nu} B_{\mu\nu}.
\end{equation}
This mixing is very undesirable for our case study, as it introduces a vector-like coupling to the quark-current, which in turn reintroduces spin-independent dark matter scattering.
For the model in \cref{app:AxialVectorModel}, such mixing is however only generated at two electroweak loops. We estimate that
\begin{equation}\label{eq:twoloopmixing}
    \epsilon \sim \frac{e^3 g'}{\sin\theta_W \cos^2\theta_W}\frac{N_c N_f}{(16\pi^2)^2} \sim 5 \times 10^{-5} \times g',
\end{equation}
with $N_c=3$ and $N_f=6$ the number of colors and flavors in the SM quark sector. 
This suppression is sufficient to guarantee that spin-dependent scattering will always dominate over spin-independent scattering in this model.

Given the constraints from the UV completion which impose \eqref{eq:thisisbad}, we can see that the direct detection cross section in the heavy mediator limit (see \eqref{eq:refcrosssecAp}) scales as 
\begin{equation}
    \sigma_{A'} \sim \left(\frac{g_{p,n}}{m_{A'}}\right)^4.
\end{equation}
Given this, we next consider constraints on the couplings and show them in terms of $g_{p/n}/m_{A'}$ in \cref{fig:Aprime_constraints}. We will then take the largest allowed value of this quantity to maximize the direct detection rate.

If light enough, the $A'$ can be produced in exotic meson decays, such as $K\to \pi A'$ or $B\to K A'$. We consider searches where the $A'$ is invisible, since the dominant decay is generally either $A' \to \bar\nu \nu $ or $A' \to\bar\chi\chi$.
Since the $A'$ is an axial vector, the two body decay $A'\to 2\pi$ is not permitted, and instead the three-body decay $A'\to 3\pi$ is the largest hadronic decay mode. 
Even if this is open, it is expected to be suppressed due to the smaller phase space and due to being a three-body decay.
Similarly, decays to e.g.~electrons through \eqref{eq:kineticmixing} are always negligible due to the smallness of $\epsilon$ in \eqref{eq:twoloopmixing}.

The strongest bounds therefore come from searches for $K\to \pi\bar\nu\nu$ and $B\to K\bar\nu\nu$, performed most recently at NA62 \cite{NA62:2021zjw} and Belle II \cite{Belle-II:2023esi}, respectively. 
The latter search has a slight excess, but in this work we do not attempt to reinterpret this result as a bound on $B\to K A'$. Instead we use the older BaBar bound \cite{BaBar:2013npw}, since even this will give very strong constraints.
 Using the computations in \cite{dror2017dark}, we find 
\begin{align}
\text{Br}&[K^\pm \to \pi^\pm A'] \approx 8.9\times 10^{-7} \left(\frac{g'}{10^{-8}}\right)^2 \left(\frac{1\,\text{MeV}}{m_X}\right)^2\nonumber\\
&\times \sqrt{\left(1-\frac{m_\pi^2}{m_K^2}-\frac{m_{A'}^2}{m_K^2}\right)^2-4 \frac{m_\pi^2 m_{A'}^2}{m_K^4}}\\
\text{Br}&[B^\pm \to K^\pm A'] \approx 2.2\times 10^{-3} \left(\frac{g'}{10^{-8}}\right)^2 \left(\frac{1\,\text{MeV}}{m_X}\right)^2\nonumber\\
&\times \mathcal{F}_K^2(m_{A'}) \sqrt{\left(1-\frac{m_K^2}{m_B^2}-\frac{m_{A'}^2}{m_B^2}\right)^2-4 \frac{m_K^2 m_{A'}^2}{m_B^4}}, 
\end{align}
with 
\begin{equation}
    \mathcal{F}_K^2(m_{A'}) \approx \frac{0.33}{1-m_{A'}^2/\mathrm{22\, GeV^2}}.
\end{equation}
The corresponding NA62 and BaBar bounds are shown in \cref{fig:Aprime_constraints}.


To our knowledge, no dedicated calculations for axial vectors have been performed in the context of supernova cooling limits. 
However, we can obtain a rough estimate of the SN1987A bound by using the Goldstone equivalence theorem, which is strictly speaking valid only for $m_{A'}\ll T$, with \mbox{$T\approx 10$ MeV} the temperature at the supernova's core.
We rescale the bounds on the axion-nucleon couplings obtained in \cite{Lella:2023bfb} for pseudoscalars by taking
\begin{equation}
g' \approx \frac{m_{A'}}{m_p} g_{ap},
\end{equation}
with $g_{ap}$ the axion-proton coupling in \cite{Lella:2023bfb}. In this formula we used mapping from the axion-quark to axion-proton coupling  derived in \cite{GrillidiCortona:2015jxo}.
We must further account for the fact that in our model the $A'$ decays to neutrinos with a width of 
\begin{equation}
    \Gamma(A'\to\bar\nu\nu) = \frac{3 g'^2 m_{A'}}{2\pi}.
\end{equation}
We require that the $A'$ is long-lived enough to exit the neutrino sphere of the supernova, which we take to be 10 km. 
This slightly modifies the constraint in \cite{Lella:2023bfb} for large couplings and masses $m_{A'}\gtrsim$ 10 MeV. This estimate leads to the dashed region shown in \cref{fig:Aprime_constraints}. 

We emphasize this is a very crude estimate of the constraint, since for $m_{A'}\gtrsim$ MeV the Goldstone equivalence theorem is not strictly valid. A more accurate treatment requires dedicated calculations of the $A'$ production and trapping rate, taking into account the axial coupling and all three degrees of freedom of the $A'$.
We do not attempt this in this work. 
All that said, as we will see below, our bounds on the direct detection cross section will not depend on this region of the parameter space.

~\cref{fig:Aprime_constraints} then makes it clear that the direct detection rate will be largest if we take $m_{A'}\gtrsim m_B$, beyond the BaBar constraint. The black star indicates our benchmark choice of $g_{p/n}/m_{A'}$ .
For completeness, we note that this benchmark choice is likely in tension with bounds on $N_{\rm eff}$ from Big Bang Nucleosynthesis: an order of magnitude estimate suggests that a coupling strength of $g_{p,n}/m_{A'}\sim 2\times 10^{-4}/$GeV likely suffices to keep the right-handed neutrinos thermalized with the SM plasma, at least until the onset of the QCD phase transition.
The $N_{\rm eff}$ bound could be circumvented either by constructing a model with Majorana neutrinos or by assuming an additional entropy dump after the QCD phase transition. However, such a late entropy dump is perhaps not particularly plausible, and poses new model building challenges related to generating the correct dark matter relic density.
That said, our goal is not to build a realistic dark matter model, but to demonstrate that spin-dependent scattering of light dark matter through an $A'$ mediator is already very strongly constrained, \emph{regardless of the cosmological history}.
This will be evident from our results in \cref{sec:Apdiscussion}.

\section{DM Scattering calculation\label{sec:scatteringcalculation}}

The calculation for DM scattering in crystals is performed in two steps. First, we match the DM-nucleon interactions in \eqref{eq:def_phi}, \eqref{eq:def_a} and \eqref{eq:def_Ap_heavy} onto an effective DM-nucleus interaction. This step is no different that what has been done for WIMP dark matter, except that we only need to retain the lowest order in the momentum expansion since we focus on sub-GeV DM. In the second step, we sum over nuclei in the crystal to obtain the DM-phonon interaction rate.

\subsection{DM -- nucleus coupling}
\label{Sec:nuc_match}

We seek to replace \eqref{eq:def_general} with an effective Hamiltonian describing the interaction of the dark matter with a nucleus. It has the form
\begin{align}
\mathcal{H}^N &= - \frac{g_\chi g_{N}}{q_0^2+m_{\rm med}^2} 
 F_{\rm med}(\mathbf{q}) \, \mathcal{O}\left(\mathbf{J}_\chi, \mathbf{q}\right)\cdot \mathbf{J}\,  e^{i\mathbf{q}\cdot \mathbf{r}},\label{eq:def_general_nuclear}
\end{align}
where now $\mathbf{J}$ is the spin operator of the nucleus. This effective Hamiltonian is a valid description as long as the scattering wavelength $\sim 1/q$ is much larger than the radius of the nucleus, meaning \mbox{$q\ll 100$ MeV} for e.g.~Si. For the range of dark matter masses we consider, this is always the case. (See \cite{Fitzpatrick:2012ix} for a detailed treatment of the finite $q$ corrections, which lead to $q$-dependent spin form factors.)

The effective coupling of the mediator to the nucleus defined as
\begin{equation}
    g_N \equiv g_{p}f_p + g_n f_n.
\end{equation}
where the $f_{p,n}$ are proportionality constants to be determined from the nuclear matrix elements, for which we follow the discussion in \cite{Engel:1992bf}. 
In \eqref{eq:refcrosssecphi}, \eqref{eq:refcrossseca} and \eqref{eq:refcrosssecAp}, we defined reference cross sections in terms of the mediator-proton coupling, so it will be therefore more convenient to define the proportionality factor
\begin{equation}
    \lambda = \frac{g_N}{g_p}= f_p + \frac{g_n}{g_p} f_n. \label{eq:deflambda}
\end{equation}
The parameter $\lambda$ therefore parametrizes the nuclear matrix elements through the $f_{p,n}$ as well as some model dependence through the ratio $g_n/g_p$. For most of the direct detection targets of interest, $f_p \gg f_n$ (see \cref{table:response_functions}). Furthermore, considering our benchmark models, we have $g_n/ g_p \lesssim 1$ for most of the parameter space of the spin-zero mediators, and $g_n/g_p = 1$ for the $A'$ mediator. So one may generally consider $\lambda \approx f_p$, though we do not use this approximation in our numerical results.


Consider a nucleus of spin $J$, with initial and final states of magnetic quantum number $m_{i,f}$. To determine the $f_{n,p}$, we need to know the nuclear matrix elements 
\begin{align}
    \langle J m_f |\mathbf{S}^{\mathrm{tot}}_p| J m_i\rangle &\equiv \langle J m_f | \sum_{i=1}^Z \mathbf{S}_p^i | J m_i \rangle \label{eq:sumspin_p}\\
    \langle J m_f |\mathbf{S}^{\mathrm{tot}}_n | J m_i \rangle &\equiv \langle J m_f | \sum_{i=1}^{A-Z} \mathbf{S}^i_n | J m_i \rangle,\label{eq:sumspin_n}
\end{align}
where $i$ labels the protons or neutrons in the nucleus. By the Wigner-Eckart theorem, expectation values of vector operators $ {\bf S}^{\mathrm{tot}}_{n/p}$ and $\mathbf{J}$ are proportional to each other with a factor which is independent on $J$ and $m_{i,f}$, {\emph{i.e.}},
\begin{equation}
    \langle J m_f | \mathbf{S}^{\mathrm{tot}}_{p/n} | J m_i \rangle = f_{p/n} \langle J m_f | \mathbf{J} | J m_i \rangle .
\end{equation}
for all $m_{i,f}$. The coefficients $f_{p/n}$ can then be determined from computing the matrix element for the $z$-component of the sum of nucleon spin operators for $m_i=m_f=J$\footnote{With another application of the Wigner-Eckart theorem, one could write $\langle J m_f | S^{\mathrm{tot},z}_{p/n} | J m_i \rangle = \frac{m_f}{\sqrt{J(J+1)}}\delta_{m_i, m_f} \frac{\langle J || S^{\mathrm{tot}}_{p/n} || J \rangle}{\sqrt{2J+1}}$
where we employ the definition of the reduced matrix element found in \cite{Sakurai_Napolitano_2020}.
 This allows us to write the coefficients as
$f_{p/n} = \frac{\langle J || S^{\mathrm{tot}}_{p/n} || J \rangle}{\sqrt{J (J+1)(2J+1)}}$ 
which is manifestly independent on the magnetic angular momentum quantum numbers.},
\begin{equation}
    f_{p/n} = \frac{\langle J J | S_{p/n}^{\mathrm{tot},z} | J J \rangle}{J}.
\end{equation}
  
The $\langle J J | S_{p/n}^{\mathrm{tot},z} | J J \rangle$ matrix elements can be determined from shell-model calculations \cite{Fitzpatrick:2012ix,Klos:2013rwa,Hu:2021awl}.  Since these calculations have not yet been performed for all materials of interest to us, we also use a phenomenological model known as the Odd-Group Model (OGM) to calculate the values of $f_{p/n}$ \cite{shalit, Engel:1992bf, PhysRevD.40.3132}.

The OGM applies to odd-mass nuclei, which contain an odd number of nucleons of one type and an even number of nucleons of the other type. In the OGM, it is assumed that the nuclear magnetic moment is entirely determined by the odd group nucleons. 
Then the spin matrix elements are estimated as  
\begin{align}
\langle J J | S_{\rm odd}^{{\rm tot}, z} | J J \rangle &\simeq \frac{\mu_N - g_{\rm odd}^l J}{g_{\rm odd}^s - g_{\rm odd}^l}\\
\langle J J |  S_{\rm even}^{{\rm tot}, z} | J J \rangle &\simeq 0,
\end{align}
where $\mu_N$ is the magnetic moment of the nucleus, as measured experimentally. The subscript ``odd'' stands for either $p$ or $n$, depending on whether there are an odd number of protons or neutrons. Neglecting meson-exchange currents in the nucleus, the $g$-factors for the proton (neutron) are given by 
\begin{align}
g_{p}^s &= 5.586  &g_{n}^s&=-3.826\\
g_{p}^l &= 1      &g_{n}^l&=0.
\end{align}

\cref{table:response_functions} shows a comparison between both approaches for a number of relevant isotopes, along with their natural abundance.
The OGM performs reasonably well and is always within a factor of 2 of the shell model predictions for the odd-group coupling, except for ${}^{127}$I.
We note that different implementations of shell-model calculations also suffer from large uncertainties for ${}^{127}$I \cite{Hu:2021awl}.


{\setlength{\tabcolsep}{5.8pt}
\begin{table*}[tb] 
\centering
\begin{tabular}{c | c c c c| c c c c| c | c}
\hline\hline
 & \multicolumn{4}{c|}{OGM~\cite{PhysRevD.40.3132}} &  \multicolumn{4}{c|}{Shell-model~\cite{Klos:2013rwa}} & Abundance & Spin\\ 
\cline{2-9}
& $f_p$ & $f_n$ & $\lambda_{a,\phi}$ & $\lambda_{A'}$ & $f_p$ & $f_n$ & $\lambda_{a,\phi}$ & $\lambda_{A'}$ & [\%] & $J$  \\
\hline
${}^{127}$I & 0.027 & 0     & 0.027   & 0.027 & 0.137 & 0.012 & 0.136  & 0.149  & 100 & 5/2\\
${}^{73}$Ge & 0     & 0.051 & -0.002  & 0.051   & 0.007 & 0.098 & 0.003 &  0.105 &   7.76 & 9/2\\
${}^{29}$Si & 0     & 0.291 &  -0.01  & 0.291  & 0.032 & 0.312 & 0.019 &  0.344 &   4.7 & 1/2\\
${}^{27}$Al & 0.100 & 0     & 0.100   & 0.100   & 0.131 & 0.015 & 0.13& 0.146  &   100 & 5/2\\
${}^{23}$Na & 0.105 & 0     &0.105   & 0.105  & 0.150 & 0.016 & 0.149 &  0.166 &   100 & 3/2\\
${}^{19}$F  & 0.929 & 0     & 0.929   & 0.929   & 0.956 & 0.004 & 0.956  &  0.960 & 100 & 1/2\\
${}^{69}$Ga & 0.075 & 0     & 0.075  & 0.075  & \_ & \_ & \_   & \_   & 60.1 & 3/2\\
${}^{71}$Ga & 0.154 & 0     & 0.154  & 0.154  & \_ & \_ & \_   & \_   & 39.9 & 3/2\\
${}^{75}$As & 0.009 & 0     & 0.009  & 0.009  & \_  & \_& \_   & \_   & 100 & 3/2\\
\hline \hline 
\end{tabular} 
\caption{ Comparison of results from the Odd-Group Model (OGM) with shell-model calculations in \cite{Klos:2013rwa} for various isotopes relevant for spin-dependent direct detection. When shell-model computations are not available, we leave the entry blank. The last columns give the natural abundance and spin of each isotope. For the $a,\phi$ mediators, the factor $\lambda_{a,\phi}$ (defined in \eqref{eq:deflambda}) was calculated in the limit where $m_{a,\phi}\ll m_\pi$.}
\label{table:response_functions}
\end{table*}}

\subsection{DM -- crystal interaction}

To find the interaction of the dark matter with the crystal, we simply sum \eqref{eq:def_general_nuclear} over all atoms in the crystal
\begin{align}
\mathcal{H}^c &= - \frac{g_\chi g_{p}}{q_0^2+m_{\rm med}^2}
 F_{\rm med}(\mathbf{q}) \sum_{\ell,d} \lambda_{\ell,d} \mathcal{O}\left(\mathbf{J}_\chi, \mathbf{q}\right)\cdot \mathbf{J}_{\ell,d}\,  e^{i\mathbf{q}\cdot \mathbf{r}_{\ell,d}}\label{eq:def_general_crystal}
\end{align}
where $\ell$ runs over the crystal cells and $d$ indexes the inequivalent positions of the atoms in the unit cell. We indexed the proportionality factor defined in \eqref{eq:deflambda} with $\ell,d$ to account for different elements and isotopes that may be present at different lattice sites.

In the Born approximation, the scattering rate is given by Fermi's Golden Rule
\begin{align}
\Gamma =& \frac{2\pi}{V} \sum_{i,f } \sum_{i_\chi,f_\chi} w_i w_{i_\chi} \int \!\!\frac{d^3 \bfq}{(2\pi)^3}|\langle f, f_\chi |\mathcal{H}^c(\bfq) | i, i_\chi \rangle |^2 \nonumber\\
&\times\delta (E_f - \omega-E_i).
\end{align}
with $V$ the target volume and $\bfq$ and $\omega$ the momentum and energy deposited by the dark matter into the crystal. 
 The sums over $i$ and $f$ represent a sum over all initial and final state crystal configurations, including the spin states of the nuclei and collective phonon excitations.
 $E_i$($E_f$) is the initial(final) energy in the crystal.  We assume the crystal is initially in a ground state, so we will set $E_i=0$ later on. The initial states are weighted by the $w_i$, which are chosen such that the spins are oriented randomly in the crystal.  
The sums over $i_\chi,f_\chi$ represent the sums over the dark matter's initial and final spin configurations. We similarly choose the weights $w_{i_\chi}$ for the initial spin configuration such that the dark matter is unpolarized.

It is helpful to rewrite the $\delta$-function in terms of its Fourier transform: 
\begin{align}
\Gamma =& \frac{1}{V}\int_{-\infty}^{+\infty}\!\!\!\! dt \int \!\!\frac{d^3 \bfq}{(2\pi)^3}\sum_{i,f }\sum_{i_\chi,f_\chi} w_i w_{i_\chi} \langle i,i_\chi |\mathcal{H}^{c\dagger} | f,f_\chi \rangle \nonumber\\
&\times\langle f,f_\chi |e^{-i E_f t} \mathcal{H}^c e^{i E_i t}| i,i_\chi \rangle  e^{i\omega t}.
\end{align}
The $|i,f\ket$ are eigenstates of the crystal Hamiltonian, so we can rewrite these phases as $e^{\pm i H t}$ and time-evolve the crystal degrees of freedom in the second expectation value (see for example \cite{Griffin:2018bjn} for more details). The sum over $f$ and $f_\chi$ can then be eliminated, as they are now an unrestricted sum over a complete set of states, giving:
\begin{equation}
\Gamma = \frac{1}{V} \! \int_{-\infty}^{+\infty}\!\!\!\!\!\!\!\! dt \, e^{i \omega t} \!\!\! \int \!\!\frac{d^3 \bfq}{(2\pi)^3} \! \sum_{i,i_\chi}  w_i w_{i_\chi} \langle i,i_\chi |\mathcal{H}^c(0)^\dagger \mathcal{H}^c(t)| i,i_\chi \rangle.
\end{equation}
Plugging in the Hamiltonian in \eqref{eq:def_general_crystal} and factorizing the matrix element into DM and crystal pieces, we can decompose the matrix element as
\small
\begin{align}
& \Gamma = \frac{1}{V} \frac{g_\chi^2g_p^2}{(q_0^2+m_{\rm med}^2)^2} \int_{-\infty}^{+\infty}\!\!\!\!\!\!\! dt \,  e^{i\omega t} \!\! \int \!\!\frac{d^3 \bfq}{(2\pi)^3} |F_{\rm med}(\mathbf{q})|^2 \times \nonumber \\
& \sum_{\alpha,\beta=x,y,z}\bigg[ \sum_{i_\chi} w_{i_\chi}\langle i_\chi | \mathcal{O}\left(\mathbf{J}_\chi, \mathbf{q}\right)^\alpha \mathcal{O}\left(\mathbf{J}_\chi, \mathbf{q}\right)^\beta | i_\chi\rangle \bigg] \times \nonumber \\
& \bigg[  \sum_{i} \! w_{i} \!\!\!\!  \sum_{\ell,d ;\ell', d' } \!\!\!\!\!  \lambda_{\ell,d}\lambda_{\ell',d'}   \langle i | J^\alpha_{\ell,d} e^{-i \bfq\cdot \bfr_{\ell,d}(0)}  \!
 J^\beta_{\ell' \!\!,d'} e^{i \bfq\cdot \bfr_{\ell' \!\!,d'}(t)} | i \rangle \bigg]. \label{eq:rate_decomposed}  
\end{align}
\normalsize
Note that here we have assumed a basis of tensor product eigenstates for the DM and the nuclear spins. One may also consider a basis of \emph{total} angular momentum eigenstates of the DM and nuclei. For a random orientation of nuclei spins and unpolarized dark matter, that is for constant weights $w_i$, the two descriptions coincide by virtue of the orthogonality of the Clebsch-Gordan coefficients.
We have also assumed that the spin degrees of freedom in the crystal do not time-evolve under the action of the Hamiltonian.
In other words, we assumed that the dynamical degrees of freedom, the phonons, do not affect the orientations of the nuclear spins. 
This should be a very good approximation in non-magnetic materials.

\begin{figure*}[t!]
    \subfloat{%
    \includegraphics[width=0.48\linewidth]{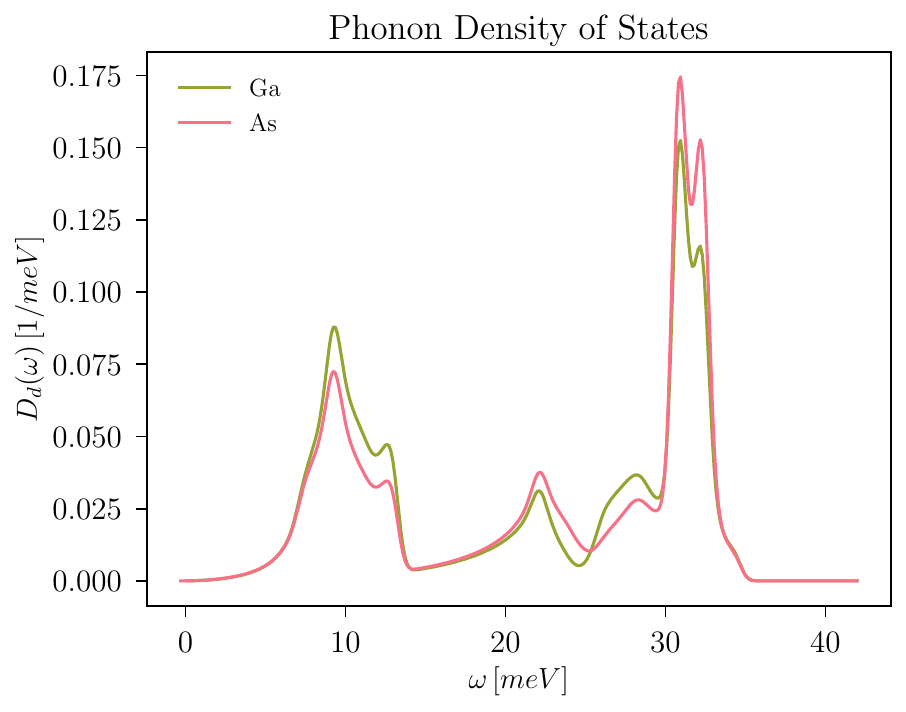}%
    }
    \subfloat{%
    \includegraphics[width=0.48\linewidth]{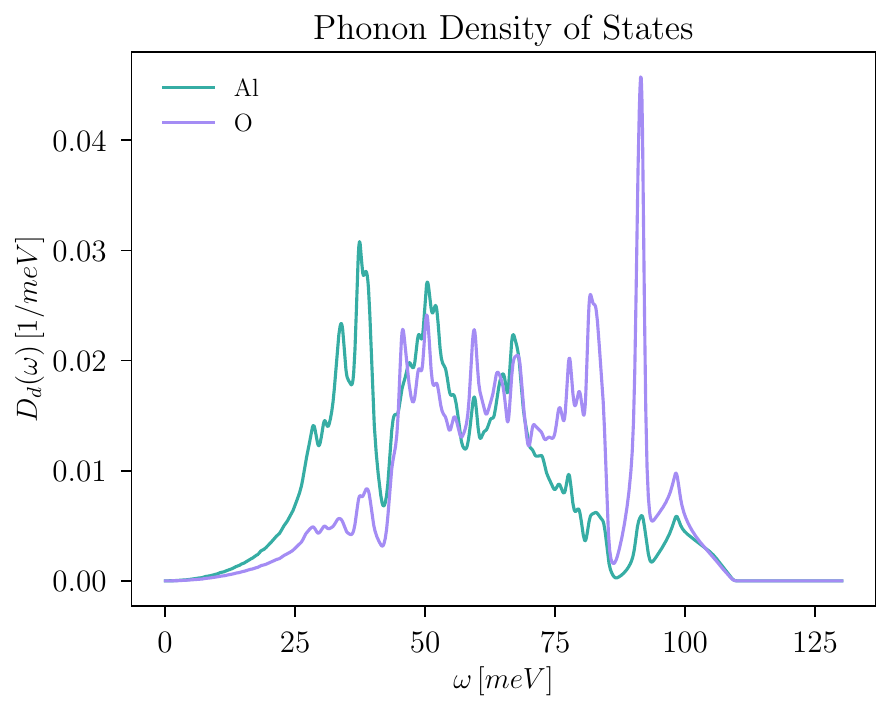}%
    }
    \caption{The phonon partial density of states for each individual atom in both GaAs and Al$_2$O$_3$ crystals \cite{schubert:2000fjv,Gervais_Piriou_1974,lawler:2004gha}.}
    \label{fig:densityofstates}%
\end{figure*}

Given that phonon degrees of freedom are decoupled from the nuclear spin states, we next factorize the crystal matrix element into two pieces. We decompose initial states as the product state of the nuclear spin configuration and the ground state of the phonon system, \mbox{$|i\rangle = | i_s \rangle \otimes |0\rangle$}, where $| i_s \rangle$ are the states corresponding to all spin configurations of the crystal and $|0\rangle$ is the (unique) zero-phonon state.  The nuclear spin matrix element can be further simplified since spin orientations are isotropically distributed. Furthermore, the cross terms with $\ell\neq\ell'$ and $d\neq d'$ average to zero when summed over all initial nuclear spin configurations. In the crystal matrix element in the last line of \eqref{eq:rate_decomposed}, we can therefore make the replacement 
\begin{align}
    \label{eq:JJ_replacement}
     J^\alpha_{\ell,d}  J^\beta_{\ell' \!\!,d'} \to \frac{1}{3}  \delta^{\alpha \beta} \delta_{\ell \ell'} \delta_{d d'} \bfJ_{\ell,d}^2. 
\end{align}

We then define the following factorized structure function which captures the crystal response:
\begin{align}
S(\bfq,\omega)\equiv&  \frac{1}{N}\sum_{\ell,d} \lambda^2_{\ell,d} C_{\ell,d}(\bfq,\omega) \sum_{i_s} w_{i_s}\langle i_s | \bfJ_{\ell,d}^2 | i_s \rangle \label{eq:structurefactor}\\
C_{\ell,d}(\bfq,\omega)\equiv &  \int_{-\infty}^{+\infty}\!\! dt \, e^{i\omega t}  \langle 0| e^{-i \bfq\cdot \bfr_{\ell,d}(0)}   e^{i \bfq\cdot \bfr_{\ell,d}(t)} | 0 \rangle,
\label{eq:Cld_def}
\end{align}
where $N$ is number of crystal unit cells (labeled by $\ell$).
For the phonon correlation function, \eqref{eq:Cld_def}, we neglect isotopic variation across crystal cells $\ell$, which gives small variations in the nuclear masses.  Then we can factor $C_{\ell, d}$ out of the sum over $\ell$, giving
\begin{align}
S(\bfq,\omega)=&   \sum_{d}\overline{\lambda^2_{d} J_d(J_d+1)} C_{\ell,d}(\bfq,\omega),\label{eq:Sqointermediate}
\end{align}
where $\overline{\lambda^2_{d} J_d(J_d+1)}$ is the average value for the atom in position $d$ over all crystal cells. This average therefore accounts for variations  in the spin-dependent coupling  across isotopes.

The $C_{\ell,d}(\bfq,\omega)$ correlation function was calculated in \cite{Campbell-Deem:2022fqm} and is given by\footnote{Relative to \cite{Campbell-Deem:2022fqm}, we defined the $C_{\ell,d}$ without the $1/V$ normalization factor, which is instead accounted for in the total rate.}
\begin{align}
C_{\ell,d}(\bfq,\omega) =& 2\pi e^{-2W_d(\bfq)}\sum_{n=1}^\infty \frac{1}{n!} \left(\frac{q^2}{2m_d}\right)^n \nonumber\\
&\times \prod^n_{i=1}\int d\omega_i \frac{D_d(\omega_i)}{\omega_i} \delta \left(\sum_{i=1}^n \omega_i -\omega\right).\label{eq:Cldresult}
\end{align}
where $D_d(\omega_i)$ is the partial phonon density of states for the cell site $d$. The $D_d(\omega)$ can be obtained using Density Functional Theory (DFT) computations, see \cref{fig:densityofstates}. $W_d(\bfq)$ is the Debye-Waller factor 
\begin{equation}
    W_d(\bfq) \equiv \frac{q^2}{4 m_d}\int d\omega'\frac{D_d(\omega')}{\omega'},
\end{equation}
with $m_d$ the mass of the atom in the unit cell indexed by $d$. The sum in \eqref{eq:Cldresult} converges to a Gaussian in the regime where $q^2/m_d$ is much larger than the typical phonon frequency \cite{Campbell-Deem:2022fqm}. This asymptotic form can be used to avoid having to calculate a large number of terms in the series.

We now write the total rate in \eqref{eq:rate_decomposed} as
\begin{align}
\Gamma=& \frac{g_\chi^2g_p^2}{(q_0^2+m_{\rm med}^2)^2}\frac{N}{V} \!\!\!  \int \!\!\frac{d^3 \bfq}{(2\pi)^3} |F_{\rm med}(\mathbf{q})|^2 G(\bfq) S(\bfq,\omega), \label{eq:Gammamaster} 
\end{align}
with $N$ the number of unit cells in the crystal and $V/N$ the volume of a single crystal unit cell.
The mediator-dependent matrix elements are captured by the quantity $G(\bfq)$: 
\begin{align}
G(\bfq)\equiv& \frac{1}{3}\sum_{i_\chi} w_{i_\chi}\langle i_\chi | \mathcal{O}\left(\mathbf{J}_\chi, \mathbf{q}\right)\cdot \mathcal{O}\left(\mathbf{J}_\chi, \mathbf{q}\right) | i_\chi\rangle.
\end{align}
For the models we consider, given in \eqref{eq:Ophi}, \eqref{eq:Oa}, and \eqref{eq:OAprime}, this is 
\begin{align}
G_\phi(\bfq) &= \frac{1}{3}\frac{|\bfq|^2}{m_p^2} \\
G_a(\bfq) &= \frac{1}{9}(J_\chi+1)J_\chi \frac{|\bfq|^4}{m_p^2m_\chi^2} \\
G_{A'}^{\rm heavy}(\bfq) &= \frac{16}{3}  (J_\chi+1)J_\chi.
\end{align}

Finally, the rate per unit of exposure is given by
\begin{equation}
    R = \frac{1}{\rho_T} \frac{\rho_\chi}{m_\chi} \int d^3 \mathbf{v} \, f(\mathbf{v}) \,\Gamma(v),
\end{equation}
with $\rho_T$ and $\rho_\chi=0.4\, \text{GeV}/\text{cm}^3$ the mass densities of the target and the dark matter, respectively. $f(v)$ is the dark matter velocity profile, which we take to be a boosted, truncated Maxwell-Boltzmann distribution with the average galaxy-frame velocity $v_0=220$ km/s, the average earth velocity $v_e =
240$ km/s, and $v_{esc} =$ 500 km/s as the local
escape velocity of the Milky Way.

The final results for all three models can be expressed in terms of the reference cross sections defined in \eqref{eq:refcrosssecphi}, \eqref{eq:refcrossseca} and \eqref{eq:refcrosssecAp}
\begin{align}
\label{eq:rate_phi}
    R_\phi &=   \frac{\bar\sigma_\phi}{\sum_d m_d} \frac{\rho_\chi}{m_\chi}  \frac{2\pi}{3} \frac{  m_p^2}{v_0^2 \mu_{\chi p}^4} \nonumber\\  
    &\hspace{-3mm} \times \int d^3 \mathbf{v} \, f(\mathbf{v})   \int \!\!\frac{d^3 \bfq}{(2\pi)^3} |F_\phi(\mathbf{q})|^2 \frac{|\bfq|^2}{m_p^2} S(\bfq,\omega)\\
    \label{eq:rate_a}
    R_a &=   \frac{\bar\sigma_a}{\sum_d m_d}\frac{\rho_\chi}{m_\chi}     \frac{\pi(J_\chi + 1) J_\chi}{3} \frac{ m_p^2 m_\chi^2}{v_0^4 \mu_{\chi p}^6}  \nonumber\\  
    &\hspace{-3mm} \times  \int d^3 \mathbf{v} \, f(\mathbf{v})   \int \!\!\frac{d^3 \bfq}{(2\pi)^3} |F_a(\mathbf{q})|^2 \frac{|\bfq|^4}{m_p^2 m_\chi^2} S(\bfq,\omega)\\
    \label{eq:rate_Aprime}
    R_{A'}^{\rm heavy} &=  \frac{\bar\sigma_{A'}^{\rm heavy}}{\sum_d m_d} \frac{\rho_\chi}{m_\chi}   \frac{16\pi(J_\chi + 1) J_\chi}{9}  \frac{1}{\mu_{\chi p}^2} \nonumber\\  
    &\hspace{-3mm} \times   \int d^3 \mathbf{v} \, f(\mathbf{v})   \int \!\!\frac{d^3 \bfq}{(2\pi)^3} |F_{A'}(\mathbf{q})|^2  S(\bfq,\omega), 
\end{align}
where we show the $J_\chi$ factors explicitly. (The expressions are however only valid for $J_\chi=1/2$, as they were derived exclusively for the models defined in \cref{sec:modelsandconstraints}.) 
The energy deposited by the DM is constrained to be $\omega = \bfq \cdot \bfv - q^2/(2 m_\chi)$. These formulas are implemented in the \textbf{DarkELF} package \cite{Knapen:2021bwg}, which is what we use to carry out our calculations. In \cref{app:darkelf}, we summarize our modifications to \textbf{DarkELF}, and also provide more detailed formulae for the rate integrals in the presence of an experimental energy threshold $E_{\rm th}$.

\section{Results\label{sec:discussion}}
We now show how the constraints discussed in \cref{sec:scalarbounds,sec:modelAp} impact the maximum achievable dark matter scattering rate in a low-threshold phonon-based experiment. We will consider experiments using the target materials Al$_2$O$_3$ (sapphire) and GaAs. Both are promising materials for near-future experiments~\cite{Essig:2022dfa,TESSERACT:2025tfw} and also have a high abundance of nuclei with spin-dependent interactions.
Following a common convention in the literature, we will plot cross sections corresponding to a rate of 3 signal events/kg-year exposure at various energy thresholds $E_{\rm th}$. We emphasize that these should \emph{not} be understood as projected sensitivities, since it is not yet known what energy thresholds, exposures and background levels will ultimately be achieved with such experiments. Instead, our cross section curves corresponding to 3 events/kg-year are only meant to give a sense of the absolute best-case sensitivity for a kg-scale experiment. 


\begin{figure*}[t!]
    \subfloat{%
    \includegraphics[width=\linewidth]{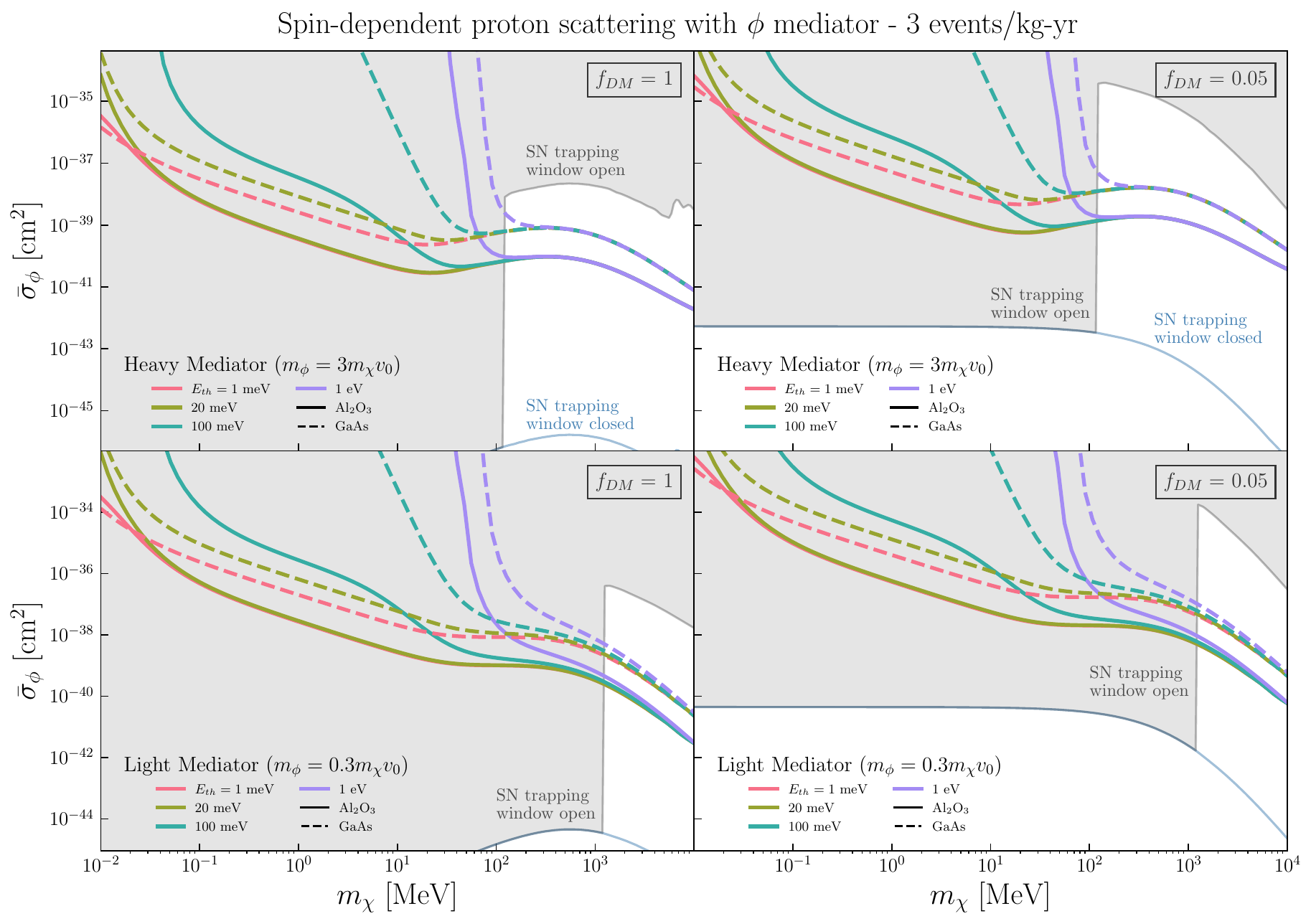}%
    }
    \caption{Reference cross sections for a signal rate of 3 events/kg-yr, through an interaction mediated by the $\phi$-mediator, in Al$_2$O$_3$ \textit{(solid)} or GaAs \textit{(dashed)} with several choices of threshold energy. 
    Existing bounds are shown by grey shaded region, and the blue line indicates where the bound would be if the supernova trapping window in \cref{fig:Constraints_med} were to close. 
    \textit{Left:} $\chi$ makes up all of dark matter ($f_{DM} = 1$) and is thus subject to the SIDM constraints in \cref{fig:Constraints_SIDM}. \textit{Right:} $\chi$ contributes only a sub-component of the total dark matter density with $f_{DM} = 0.05$. The SIDM bounds do not apply and we set $g_\chi = 1$ to satisfy the perturbative unitarity condition.
    }
    \label{fig:phi_cross}%
\end{figure*}

\subsection{$\phi$ and $a$ Mediators\label{sec:phidiscussion}} 

To present our results for the spin-0 mediators, we will use the two benchmark scenarios introduced in \cref{sec:modelsandconstraints}:
\begin{align}
    m_{a,\phi} &=0.3 \times q_0,\; \text{(light mediator benchmark)}
\label{eq:benchmarkvalueslight2}\\
    m_{a,\phi} &=3 \times q_0,\; \text{(heavy mediator benchmark)}\label{eq:benchmarkvaluesheavy2}
\end{align}
where $q_0 = m_\chi v_0$.
As a function of $m_\chi$ (and therefore $m_{\phi, a}$), we select the largest allowed $g_p$ in \cref{fig:Constraints_med} and the largest allowed $g_\chi$ in \cref{fig:Constraints_SIDM} to obtain an upper bound on $\bar \sigma_{\phi, a}$.
A priori, one could get a somewhat weaker bound by also freely floating $m_{\phi,a}$ when maximizing the direct detection cross section, subject to the bounds in \cref{sec:scalarbounds}.  
The choices \eqref{eq:benchmarkvalueslight2} and \eqref{eq:benchmarkvaluesheavy2} will however allow for a cleaner interpretation of the results; we defer the more general case to \cref{app:MoreResults}.
The qualitative conclusions are the same for both cases. 


\begin{figure*}[t!]
    \subfloat{%
    \includegraphics[width=\linewidth]{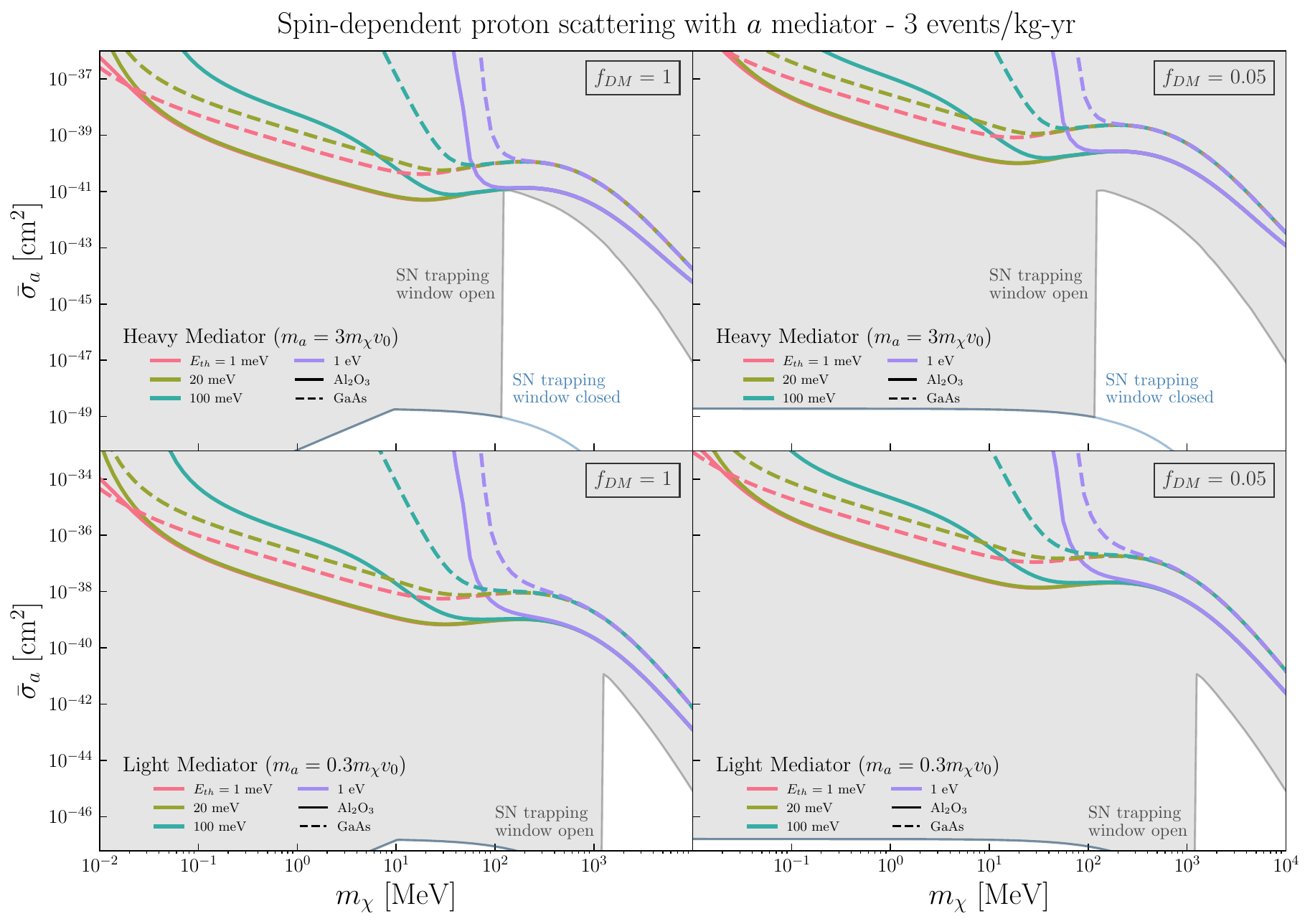}%
    }
    
    \caption{Reference cross sections for a signal rate of 3 events/kg${\text -}$yr, through an interaction mediated by the $a$-mediator, in Al$_2$O$_3$ \textit{(solid)} or GaAs \textit{(dashed)} with several choices of threshold energy.  See the caption of \cref{fig:phi_cross} for further details.}
    \label{fig:a_grids_cross}
\end{figure*}

The bounds resulting from this procedure are shown in \cref{fig:phi_cross} and \cref{fig:a_grids_cross} (grey shading), along with the cross sections for 3 events/kg-yr at various energy thresholds in sapphire (solid lines) and GaAs (dashed lines).
The left-hand panels assume that $\chi$ makes up all of the dark matter density ($f_{\rm DM} = 1$), in which case we take $g_\chi$ to saturate the self-interacting dark matter bounds discussed in \cref{sec:SIDMConstraints}.
The right-hand panels assume that $\chi$ makes up only a small fraction of the total dark matter density ($f_{\rm DM}=0.05$). 
If the remaining 95\% of the dark matter density is collisionless, strong self-interactions for the $\chi$-particle might not be sufficient to modify the dynamics of dark matter halos in a manner that would have been observed already.
In this case, $g_\chi$ is therefore only bounded by perturbative unitarity and we fix $g_\chi = 1$. 

We find that the existing bounds on $\bar \sigma_{\phi, a}$ are very strong when compared to the direct detection prospects. For the $a$-mediator, there is no viable parameter space, even with very optimistic assumptions about the detectable direct detection rate.
For the $\phi$ mediator, there is  viable parameter space for $m_\chi \gtrsim 100$ MeV ($m_\chi \gtrsim 1$ GeV) for the heavy mediator (light mediator) benchmark. In these cases, a minimal exposure of roughly a few g-day would be needed to probe new parameter space. 

Given our choice of the light and heavy benchmarks, \eqref{eq:benchmarkvalueslight2} and \eqref{eq:benchmarkvaluesheavy2}, it is straightforward to understand the dramatic edges in the cross section constraints in \cref{fig:phi_cross} and \cref{fig:a_grids_cross}. These edges occur when the constraint on $g_p$ is forced below the supernova bounds in \cref{fig:Constraints_med}. This occurs when $m_{\phi, a}$ reaches about $\sim$ 0.3 MeV.  
For \mbox{$m_{\phi, a} \gtrsim 0.3 $ MeV}, the optimal value is $g_p \approx 7 \times 10^{-6}$, in the trapping window.
However, once $m_{\phi, a} \lesssim 0.3 $ MeV, this trapping window is filled in by constraints from horizontal branch stars and the largest allowed value is therefore below the supernova bound, $g_p \approx 6 \times 10^{-10}$. 

This brings us to an important subtlety: in the case of the $\phi$ mediator, 
the open parameter space to the right of aforementioned edge is exploiting the trapping window, or the gap between the meson constraints and the SN1987A bound in \cref{fig:Constraints_med}.
To illustrate the importance of this gap, in \cref{fig:phi_cross} and \cref{fig:a_grids_cross} we show as blue lines the bounds on the cross section if the trapping window were entirely closed. In this scenario, the bounds on $g_{p,n}$ would become much more stringent, as the only viable parameter space would be below the supernova bound in \cref{fig:Constraints_med}. In this case, there would be no prospects for direct detection in any of the cases discussed above, even under the most optimistic experimental conditions.


Before moving on the $A'$-mediator, we briefly comment on some notable features in \cref{fig:phi_cross} and \cref{fig:a_grids_cross}.
The signal rate for Al$_2$O$_3$ is generally higher than for GaAs, since Al$_2$O$_3$ has similar values for the spin-dependent coupling $\lambda_{\phi, a}$ as GaAs (see \cref{table:response_functions}) but contains more nuclei per target mass due to the lower atomic mass. For specific energy thresholds, such as $E_{th} = 100$ meV, there is also a large difference in the rate due to differences in the phonon density of states of the two materials (see \cref{fig:densityofstates}).

The cross section curves for $E_{th} < 1$ eV have a notably different shape compared to the $E_{th} = 1$ eV curve.
This is because for $m_\chi\lesssim 100$ MeV the physics is described by multiphonon excitations, while the nuclear recoil approximation is sufficient for $m_\chi\gtrsim 100$ MeV ~\cite{Campbell-Deem:2022fqm}.
In addition, the downward trend of the cross section for $m_\chi > 1$ GeV may be counter-intuitive to readers familiar with spin-dependent scattering of WIMPs. 
This is because of the momentum factors in the numerators of the Hamiltonians in \eqref{eq:def_phi} and \eqref{eq:def_a}, as compared to \eqref{eq:def_Ap_heavy}. This leads to an increase in rate with increasing $m_\chi$, as long as $m_\chi\ll m_N$. This can be seen most easily from the momentum integrals in \eqref{eq:rate_phi} and \eqref{eq:rate_a}. For dark matter that is heavier than the nuclei in the crystal ($m_\chi\gg m_N$), the scattering rate scales as $\sim 1/m_\chi$.


Finally, we note that we did not attempt to explain or predict the dark matter's relic density.
In other words, we have implicitly assumed that the relic density is set by another mechanism, which is not related to the mediators that are needed to observe $\chi$ in direct detection experiments.
Accounting for a such a mechanism may lead to further bounds, especially from the BBN and/or CMB observations. 
For example, as it freezes out, the dark matter may dump additional entropy into the mediator, which could put the model in tension with the bounds on the number of relativistic degrees of freedom in the early universe \cite{Dror:2023fyd}. See \cref{app:MoreResults} for a more extended discussion.

\subsection{$A'$ Mediator \label{sec:Apdiscussion}}

\begin{figure}[t!]
    \subfloat{%
    \includegraphics[width=0.99\linewidth]{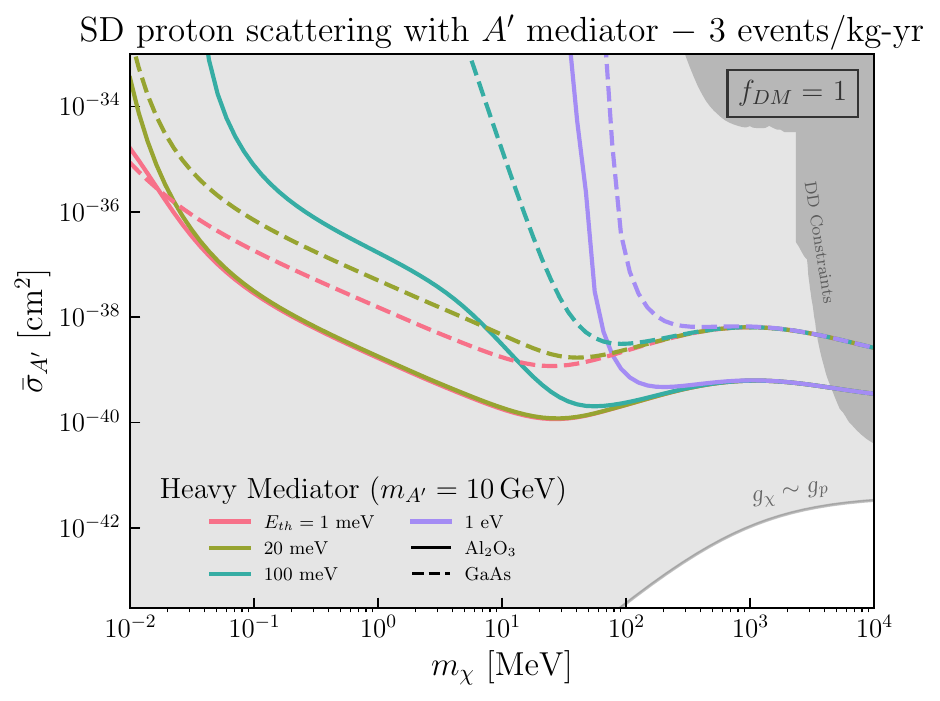}%
    \label{subfig:Aprime_hm_fDM1}}
    \caption{Reference cross sections for a signal rate of 3 events / kg${\text -}$yr, through an interaction mediated by the $A'$-mediator, in Al$_2$O$_3$ \textit{(solid)} or GaAs \textit{(dashed)} with several choices of threshold energy. The light grey shaded region denotes the bounds on $g_p$ shown in \cref{fig:Aprime_constraints}, combined with the constraint that $g_\chi \sim g_p$ for a self-consistent UV completion. The dark shaded region shows the combined existing direct detection constraints~\cite{Behnke:2016lsk,SuperCDMS:2017nns,PICO:2017tgi,CRESST:2022dtl}. We assume $\chi$ is all of dark matter ($f_{DM} = 1$) and do not consider sub-component DM since, in this case, SIDM bounds do not have constraining power.  
    }
    \label{fig:Aprimes_cross}
\end{figure}

As we showed in \cref{sec:modelAp}, there exist very stringent bounds on the $A'$ mediator, including the requirement that $g_\chi \sim g_p$ for a self-consistent theory.
These bounds are especially strong for the light mediator limit, and as a result all direct detection prospects are ruled out by many orders of magnitude. 
We therefore focus our discussion just on the heavy mediator regime.
In \cref{fig:Aprimes_cross}, we show the existing bounds on the heavy mediator $A'$ model, as compared to the cross sections needed for a rate of 3 events/kg-year. 
We see that there are likely no direct detection prospects for spin-dependent scattering within the $A'$ model, even though the operator in \eqref{eq:OAprime} is not suppressed by factors of $\bfq$.
If the condition $g_\chi \sim g_p$ could be relaxed, there is potential for viable parameter space, but constructing consistent models which achieve this is very challenging (see \cref{app:AxialVectorModel} and \cref{app:doubleaxialvectormodel}).

The condition $g_\chi \sim g_p$ implies that the direct detection cross section is proportional to the combination $\sim g_p^4/m_{A'}^4$.
The parameter choice that maximizes this combination, while consistent with existing bounds, is $m_{A'}=$10 GeV and $g_p = 1.5 \times 10^{-3}$ (see \cref{fig:Aprime_constraints}).  
 The bound in this regime comes from the LHC bounds on the heavy colored particles that are needed to cancel the gauge anomalies. 
These bounds are somewhat model-dependent, because they depend on the decay modes one assumes for those heavy colored particles (see \cref{app:AxialVectorModel}).
It is likely possible to weaken the bound on $\bar \sigma_{A'}$ by about an order of magnitude, by constructing a custom model where the colored fermions can still be hidden at the LHC, e.g.~by engineering a final state with a high number of jets.  
However, even in this scenario, the prospects for detecting low mass dark matter through spin-dependent scattering remain limited. 

For $m_\chi\gtrsim10$ GeV, experiments such as PICO \cite{PICO:2017tgi} currently do probe cross sections that are comparable to the existing bounds on the $A'$-mediator.
In other words, if a hint of dark matter in this regime were to be found in spin-dependent scattering, it would be important to scrutinize the LHC data for signs of heavy colored particles that still could have evaded the current searches.

\section{Conclusions \label{sec:summary}}


Experiments aiming to detect multiple and
single phonon interactions are coming online and promise to probe lower DM
masses ($m_\chi\lesssim 10$ GeV) than current direct detection experiments. The models that predict primarily spin-dependent DM interactions are highly constrained by searches for the mediator particle, and, in particular, by meson decays, stellar cooling, and cosmology. 
However, some regions of parameter space remain unprobed and they can lead to signals at direct detection experiments. Our main findings are:
\begin{itemize}
    \item Typically, direct detection constraints on spin-dependent scattering have assumed an operator $\propto {\bf J}_\chi \cdot {\bf S}_p$. We consider the simplest UV-complete model that achieves this with an axial vector mediator $A'$. For \mbox{$m_\chi \lesssim$ 10 GeV}, the upper bounds on direct detection cross sections 
    are extremely strong and beyond the reach of experiments that are currently being considered. 
    \item For the $a$ (axion-like particle) mediator, direct detection is both spin-dependent and momentum suppressed. If light enough, the $a$ can be produced in rare meson decays, which places strong upper bounds on the cross section. These are out of reach even for a zero-background kg-year experiment. 
    
    \item The $\phi$ mediator with mixed couplings of $\phi \bar \chi \chi$ and $g_p \bar p \gamma^5 p$ has open parameter space, which critically relies on the trapping window between the meson and supernovae bounds. 
    In the open parameter space, the maximal signal rate in $\text{Al}_2\text{O}_3$ is of the order of a hundred events per g-yr exposure.
\end{itemize}

One can construct other models and operators beyond the ones we have considered here \cite{Fitzpatrick:2012ix,Trickle:2020oki}. They all involve operators of higher dimensions and therefore additional factors of the momentum transfer, which further suppress the scattering rate. 
This is because one effectively expands in $q/m_N \sim m_\chi v / m_N$, which is always much smaller than 1 for sub-GeV dark matter. 
On the other hand, accelerator experiments place bounds on the mediators in the relativistic regime, and are therefore not bound by the same suppression factors. 
While strictly speaking we cannot rule out that a convoluted counter-example could be constructed, we believe our qualitative conclusions about spin-dependent scattering for light dark matter are broadly applicable, regardless of the choice of model.   

One exception to this argument are models with a direct coupling between the magnetic dipole moment of the DM and that of the nucleus. This interaction appears at the same order in $q$ as those discussed above but requires the DM to have a dipole coupling under the SM photon. This removes the need to introduce a new mediator. However, this model also has spin-independent dipole-charge interactions that dominate over the spin-dependent coupling to nuclei. Furthermore, as noted in \cite{Trickle:2020oki,Berlin:2025uka}, the dipole moment of the nucleus is always suppressed by a factor of $m_e/m_N$ compared to the dipole moment of the electron and thus the spin-dependent interactions of this model would be better probed via an electron recoil experiment.


\begin{acknowledgments}

We thank Reuven Balkin, Itay Bloch, Jae Hyeok Chang, Amalia Madden, Giacomo Marocco, Bashi Mandava, Samuel D. McDermott, Shijun Sun, Wolfgang Altmannshofer, Pierce Giffin, Pedro Guillaumon and Kevin Zhou for helpful discussions.
The work of BS is supported by the NSF GRFP Fellowship. 
The work of SK is supported by the Office of High Energy Physics
of the U.S. Department of Energy under contract DEAC02-05CH11231.
The research of SG and PM is supported in part by the U.S. Department of Energy grant number DE-SC0010107. TL was supported by the US Department of Energy Office of Science under Award No. DE-SC0022104 and a Harold and Suzy Ticho Endowed Fellowship.
This material is based upon work supported by the National Science Foundation Graduate Research Fellowship Program under Grant No.\ DGE 2146752. Any opinions, findings, and conclusions or recommendations expressed in this material are those of the author(s) and do not necessarily reflect the views of the National Science Foundation.
This work was initiated and performed in part at Aspen Center for Physics, which is supported by National Science
Foundation grant PHY-2210452 and the Simons Foundation (1161654, Troyer).

\end{acknowledgments}

\appendix

\section{Additional Results}
\label{app:MoreResults}

For spin-0 mediators, we previously chose benchmarks where we fixed the dark matter - mediator mass ratio: $m_{a,\phi} = 3 m_\chi v_0$ for a heavy mediator and $m_{a,\phi} = 0.3 m_\chi v_0$ for a light mediator. This somewhat arbitrary condition can be relaxed, such that a heavy (light) mediator would correspond to any $m_{a,\phi} \geq 3 m_{a,\chi} v_0$ ($m_\phi \leq 0.3 m_\chi v_0$). One can then open up the most parameter space by marginalizing over all possible mediator masses for each point in $m_\chi$ and choosing the mediator mass that produces the weakest existing constraint on the reference cross section. 

\begin{figure*}[t!]
    \subfloat{%
    \includegraphics[width=\linewidth]{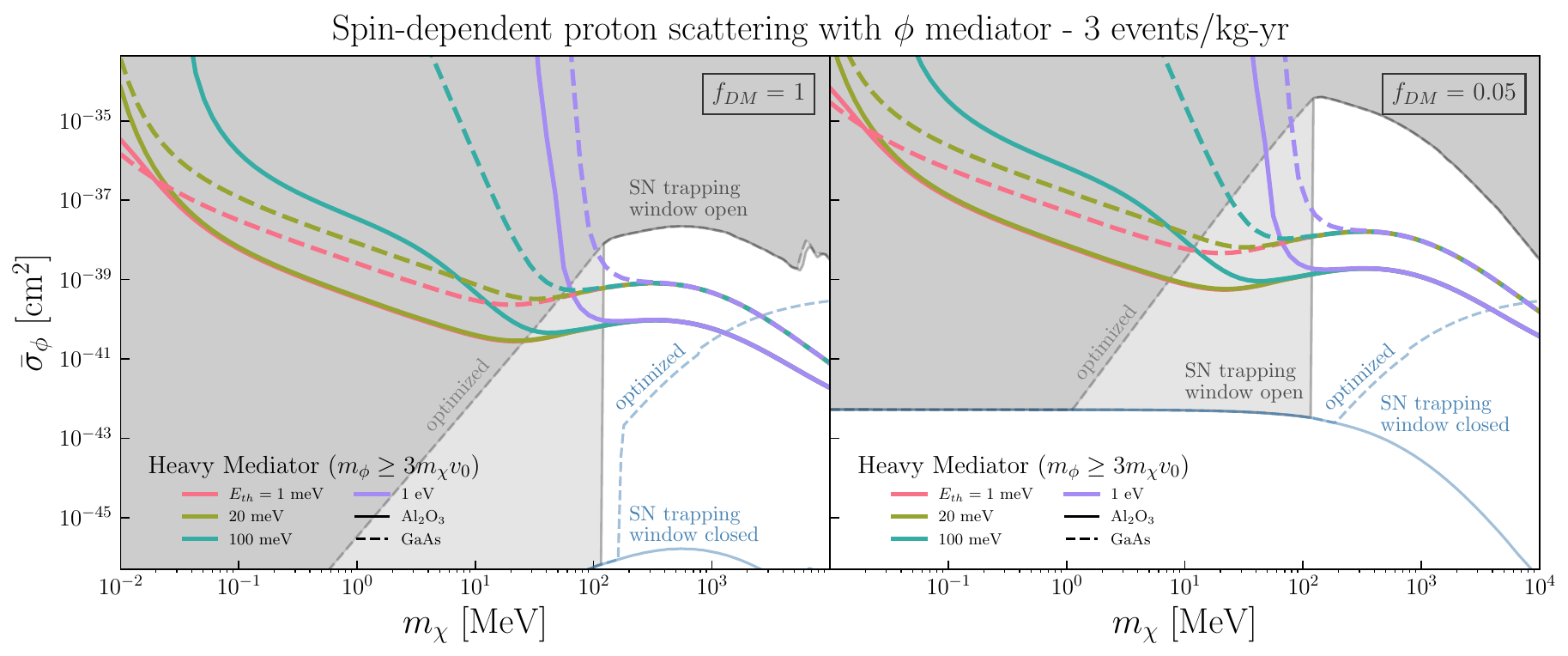}%
    }
    \caption{
    Reference cross sections for a signal rate of 3 events/kg-yr, through an interaction mediated by the $\phi$-mediator, in Al$_2$O$_3$ \textit{(solid)} or GaAs \textit{(dashed)} with several choices of threshold energy. See the caption of \cref{fig:phi_cross} for the definition of the plots and lines. The solid gray and blue lines match those shown in \cref{fig:phi_cross}. The new additions are the dashed gray and blue lines which show the bounds where $m_\phi$ is optimized for each point in $m_\chi$ to maximize the allowed cross section. For the heavy mediator, the optimized $m_\phi$ must fulfill the condition that $m_\phi \geq 3 m_\chi v_0$. For the light mediator, when the condition is relaxed to $m_\phi \leq 0.3 m_\chi v_0$, our benchmark is already optimized and therefore we do not show it here. }
    \label{fig:phi_cross_opt}%
\end{figure*}

\begin{figure*}[t!]
    \subfloat{%
    \includegraphics[width=\linewidth]{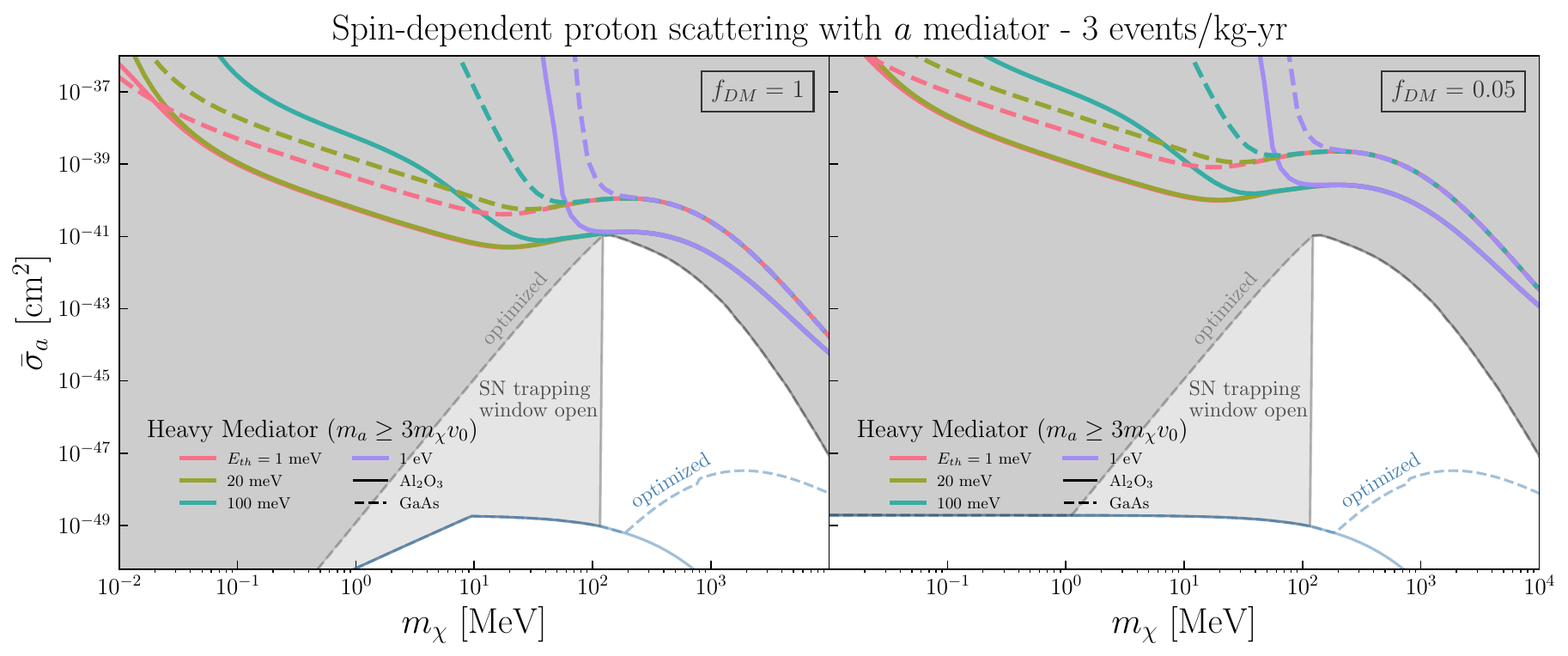}%
    }
    \caption{Reference cross sections for a signal rate of 3 events/kg-yr, through an interaction mediated by the $a$-mediator, in Al$_2$O$_3$ \textit{(solid)} or GaAs \textit{(dashed)} with several choices of threshold energy.  See the caption of \cref{fig:phi_cross} for the definition of the plots and lines. The solid gray and blue lines match those shown in \cref{fig:a_grids_cross}. 
    The new additions are the dashed gray and blue lines which show the bounds where $m_a$ is optimized for each point in $m_\chi$ to maximize the allowed cross section. For the heavy mediator, the optimized $m_a$ must fulfill the condition that $m_a \geq 3 m_\chi v_0$. For the light mediator, when the condition is relaxed to $m_a \leq 0.3 m_\chi v_0$, our benchmark is already optimized and therefore we do not show it here. }
    \label{fig:a_cross_opt}%
\end{figure*}

In \cref{fig:phi_cross_opt,fig:a_cross_opt}, these optimized bounds are shown in comparison to our default heavy benchmark bounds. The light benchmark $m_{a,\phi} = 0.3 m_\chi v_0$ presented in \cref{sec:phidiscussion} already optimizes the parameter space, and we therefore do not show  new plots for the light mediator case. The new exclusion curves are denoted by dashed grey and blue lines, corresponding to the bounds assuming the SN trapping window is open or closed, respectively. The left panels of \cref{fig:phi_cross_opt,fig:a_cross_opt} assume $f_{DM} = 1$ and therefore the optimization process considers both the SIDM constraint on $g_\chi$ as well as the various experimental constraints on $g_p$. On the right, we assume a small fraction of DM $(f_{DM} = 0.05)$ which relaxes the SIDM constraints. For these plots, we set $g_\chi = 1$ and optimize over the experimental constraints on $g_p$.

For the heavy mediator, when the SN trapping window is open, a new wedge of parameter space is opened for $m_\chi \lesssim 100$ MeV by fixing the mediator mass at about 0.3 MeV, above the horizontal branch constraints and inside the trapping window. In the case where the SN trapping window is closed and $f_{DM} = 1$, additional parameter space is opened by allowing for heavier mediator masses which allows for a substantially larger $g_p$ and $g_\chi$. In this case, the cross section bound is weakened by considering $m_{a,\phi} \sim 100$ MeV and below the accelerator bounds.

The separation here into heavy and light mediators, with somewhat arbitrary cutoffs,  may leave one wondering whether there is open parameter space in the intermediate range of mediator masses $0.3 m_\chi v_0 < m_{\phi,a} < 3 m_\chi v_0$. We verified that the amount of open parameter space does not shift significantly based on this choice.

While the above optimization appears to open up more parameter space for MeV-scale dark matter, the mass scales and couplings here could face additional constraints from cosmology.
In fact, for these couplings, $\chi$ scattering leads to $\chi$ thermalization, likely resulting in tension with the $N_{\rm eff}$ measurement from BBN. 
To avoid $\chi$ thermalization, the scattering rate of $\chi\chi \rightarrow \{a,\phi\}\{a,\phi\}$ must be less than the Hubble scale when neutrinos decouple, forcing $g_\chi$ to be very small. Again, in this scenario, all parameter space opened by the optimization is ruled out. 
Alternatively, one may assume a very low reheating scale, such that the DM was never in thermal equilibrium. This case is the most optimistic and allows $g_p$ to live in the trapping window,  opening up significant parameter space. However, attaining the correct DM relic density with such a low reheating scale is a nontrivial task and should not be assumed lightly.

In conclusion, while carefully selecting the mediator mass can open up some potential parameter space, it is important to consider the cosmological implications of such a choice. Much of the parameter space in the new wedge likely requires non-trivial model building to avoid tension with the $N_{\rm eff}$ measurement from BBN. 

\
\section{Caveats related to CHARM bound \label{app:charm}}
We note several caveats in interpreting recent recasts of the CHARM experiment bound on the ALP parameter space \cite{Jerhot:2022chi, Berger:2024xqk, Afik:2023mhj}. First, the CHARM collaboration used two estimators for the reconstruction and selection of events from axion decays inside the detector volume~\cite{CHARM:1985anb}. 
The first estimator is a  measure of the deviation of the shower axis from the beam direction and the second estimator is a measure of the transverse width of the shower, both of which are highly sensitive to the boost of the axion, leading to a fiducial efficiency that depends on the axion mass. 
These estimators have only been mentioned briefly in the CHARM paper~\cite{CHARM:1985anb}, such that modeling them reliably at this point in time may not be possible.
In all reinterpretations that we know of, a constant fiducial efficiency has been applied, and the mass dependence of the efficiency has not been taken into account in the literature. It is not clear how valid this approximation is.

Second, fermion contributions (generated by RGE running) to the kaon decay width into an axion generally have been neglected. 
We find that the RG-generated off-diagonal down-strange-ALP coupling 
dominates the contribution from the $c_{GG}$ coupling in the amplitude of $K_L\to\pi^0 a$,
 by at least an order of magnitude. 
 This may lead to an $O(1)$ correction to the constraints quoted in the literature.

\section{Axial vector model UV completion\label{app:AxialVectorModel}}

As a proof-of-concept, we construct a UV-complete model of an axial vector with couplings to SM baryons and suppressed couplings to the charged leptons, as in \eqref{eq:anomalouslowE}.  In this model, the coupling of the dark matter to the $A'$ is predicted to be half that of the SM quarks to the $A'$.
In \cref{app:doubleaxialvectormodel}, we will attempt to construct a more involved model in which this prediction can be relaxed.

\begin{table}[t]
    \centering
    \begin{tabular}{c|cccc}
         & SU(3) & SU(2) & $U(1)_Y$ & $U(1)'$\\\hline
       $Q$  & $\square$ & $\square$ & 1/6 & +1\\
       $u$  & $\overline\square$ & $1$ & -2/3 & +1\\
       $d$  & $\overline\square$ & $1$ & 1/3 & +1\\
       $L$  & 1 & $\square$ & -1/2 & 0\\
       $e$  & 1 & 1 & +1 & 0\\
       $N$  & 1 & 1 & 0 & +2\\
       $H$ & 1 &$\square$ & -1/2 & 0\\\hline
       $Q'$  & $\square$ & $\square$ & 1/6 & -1\\
       $u'$  & $\overline\square$ & $1$ & -2/3 & -1\\
       $d'$  & $\overline\square$ & $1$ & 1/3 & -1\\
       $\bar Q'$  & $\overline\square$ & $\square$ & -1/6 & 0\\
       $\bar u'$  & $\square$ & $1$ & 2/3 & 0\\
       $\bar d'$  & $\square$ & $1$ & -1/3 & 0\\
       $ N'$  & 1 & 1 & 0 & -2\\
       $\bar N'$  & 1 & 1 & 0 & 0\\
       $H_u^\prime$ & 1 & $\square$ & 1/2 &-2 \\
       $H_d^\prime$ & 1 & $\square$ & -1/2 &-2 \\
       $\phi$ &1 &1 & 0 &$+1$\\\hline
       $\chi_1$ &1 &1 & 0 &$-1/2$\\
       $\chi_2$ &1 &1 & 0 &$-1/2$\\
       $\bar\chi_1$ &1 &1 & 0 &$1/2$\\
       $\bar\chi_2$ &1 &1 & 0 &$1/2$
    \end{tabular}
    \caption{Matter content for an example UV completion of the axial $A'$ model. All fermion fields are chosen to be left-handed Weyl fermions by convention. The second block contains a set of heavy fermions, designed to cancel the gauge anomalies. $\phi$ is a scalar field which generates the masses of the anomalons, the $A'$ and the dark matter. The third block contains the dark matter and a set of heavier dark fermions which cancel the dark matter's contribution the $U(1)'$ gauge anomalies.
    }
    \label{tab:Apmodelcontent}
\end{table}

The model is based on a new $U(1)^\prime$ gauge symmetry; the full matter content is shown in \cref{tab:Apmodelcontent}. 
We assume for simplicity that the $A'$ couples universally to all three quark generations, and assume Dirac neutrinos with sterile components $N$. The $N$ field is also charged with respect to $A'$.

We must include a set of heavy fermions (see e.g.~\cite{Ismail:2016tod}) to cancel the gauge anomalies, a scalar field $\phi$ which breaks the $U(1)'$ spontaneously, and two new Higgs inert doublets $H_u^\prime$ and $H_d^\prime$, which are responsible for generating the SM  Yukawa interactions. These are given in the second block of \cref{tab:Apmodelcontent}. The heavy fermions are vector-like relative to the SM gauge groups, and as such the SM gauge anomalies continue to cancel trivially. 
The theory is now also vector-like relative to the $U(1)'$ charges, meaning that the only non-trivial anomalies are the mixed $U(1)'$-SM anomalies, which cancel as well.
The mass terms for the anomalon fields are generated by the interactions
\begin{align}\label{eq:anomalonmasses}
\mathcal{L} \supset& \lambda_Q \phi Q'\bar Q' + \lambda_u \phi^\dagger u'\bar u' + \lambda_d \phi^\dagger d'\bar d' +\frac{\phi^2}{\Lambda_N}  N'\bar N'.
\end{align}
The last term in \eqref{eq:anomalonmasses} has operator dimension 5, but can be UV completed easily. The $N'$ and $\bar N'$ also play no role in the phenomenology, other than anomaly cancellation. They can therefore be significantly lighter than the remaining fermions in \eqref{eq:anomalonmasses} without running into existing bounds of any sort.  

The LHC bounds on the colored fermions $Q',\bar Q'$, etc., depend on their assumed decay and production channels, as reviewed in \cite{CMS:2024bni}.
If we consider only hadronic decays, the current bounds on a single generation vector-like fermion range between 800 GeV and 1400 GeV \cite{CMS:2019eqb}, depending on the assummed branching ratios and representation under the SM gauge group. 
We therefore require that their masses $\gtrsim 1$ TeV for all twelve vector-like quarks, which we consider to be conservative. 
We further assume $\lambda_{Q,u,d}\lesssim 1$, to avoid excessive fine-tuning and Landau poles at a low energy scale.
This is because the field $\phi$ couples to all twelve Dirac fermions in \eqref{eq:anomalonmasses} which carry three colors each. This leads to a large correction to $\phi$'s potential, even for $\lambda_{Q,u,d}\approx 1$ and a very low UV cutoff scale.
With these assumptions, we  find that we need to impose \mbox{$\langle\phi\rangle\gtrsim 1$ TeV} in \eqref{eq:anomalonmasses} to satisfy the LHC bounds on the anomalon fields. 
The field $\phi$ also generates the mass of the $A'$, such that we can obtain the bound 
\begin{equation}\label{eq:anomalonboundapp}
g' \lesssim \frac{1}{\sqrt{2}}\frac{m_{A'}}{\mathrm{TeV}}.
\end{equation}
We note that $\bar\sigma_{A'} \sim g'^4$, which means that the bound shown in \cref{fig:Aprimes_cross} is rather sensitive to the assumptions we make about the UV theory. 
The bound on $\bar\sigma_{A'}$ could therefore easily be weaker by an order of magnitude or more if one  assumes more aggressive values for either $\lambda_{Q,u,d}$ or if the LHC bounds can be relaxed, e.g.~by assuming a final state with a large number of soft jets.
Still, this would not affect the conclusions we have drawn in \cref{sec:Apdiscussion}.

Next, we must generate the Yukawa couplings for quarks and leptons with the SM Higgs, $H$. 
This can be achieved using the $H_{u,d}^\prime$ interactions 
\cite{Kahn:2016vjr}\begin{align}
\mathcal{L}\supset&\,
y_e H L e + Y_U (H_u^\prime)^{c\dagger} Q u + Y_D H_d^\prime Qd + Y_E (H_u^\prime)^{c\dagger} LN \nonumber \\ & + \tilde\lambda_u \phi^2 H_u^\prime H + \tilde\lambda_d \phi^2 H_d^\prime H^\dagger- m^2_{H'_u}|H'_u|^2 - m^2_{H'_d} |H'_d|^2 \nonumber \\ &+ \rm{h.c.}
\end{align}
where $H_{u,d}^c\equiv i\sigma_2 H_{u,d}^*$. Upon integrating out the heavy fields $H_{u,d}^\prime$, and spontaneously breaking the $U(1)'$ by allowing $\phi$ to pick up  a vev\footnote{For simplicity, we consider the case where $\langle H_{u,d}'\rangle=0$.}, these terms in the Lagrangian lead to the Yukawa interactions
\begin{equation}
    \mathcal{L} \supset  y_u Q (H^c) u + y_d Q H d + y_N L (H^c) N + \rm{h.c.}
\end{equation}
where 
\begin{equation}
    y_{u,d} \sim Y_{U,D} \tilde\lambda_{u,d}\frac{\langle \phi\rangle^2}{m^2_{H_{u,d}^\prime}},~~~y_{N} \sim Y_{E} \tilde\lambda_{u}\frac{\langle \phi\rangle^2}{m^2_{H_{u}^\prime}}.
\end{equation}

For the dark sector, we use the UV completion put forward in \cite{Agrawal:2020lea}. In particular, to cancel the gauge anomalies we need a pair of Weyl fermions $\chi_{1,2}$ with charge $-1/2$ and along with a pair $\bar \chi_{1,2}$ with charge $1/2$, given in the third block of \cref{tab:Apmodelcontent}.
The dark sector Yukawa interactions are
\begin{align}\label{eq:Apdmmassterms}
    y_\chi \phi \chi_1\chi_2 + \bar y_\chi \phi^\dagger \bar\chi_1\bar \chi_2 ~ + \rm{h.c.}
\end{align}
The vectorlike mass terms of the form $\chi_{1,2}\bar\chi_{1,2}$ are assumed to be absent or heavily suppressed. 
This is needed to suppress the coupling of the $A'$ to the dark matter vector-current.
We identify the Dirac fermion composed out of the Weyl fermions $\chi \equiv (\chi_1,\chi_2^\dagger)$ with the dark matter.
We further assume that $\bar y_\chi> y_\chi$, such that the Dirac fermion associated with $\bar \chi_1$ and $\bar \chi_2$ is always heavier than $\chi$.
Going forward, we will assume that the annihilation rate of this heavier fermion to either $\chi$ or $A'$ is efficient enough that we can neglect its residual relic density.
The mass parameters of the dark sector particles are given by
\begin{align}
m_{A'}^2 &= 2 g'^2 \langle\phi\rangle^2 \\
m_{\phi}^2 &=2\lambda \langle\phi\rangle^2\\
m_\chi &= y_\chi \langle\phi\rangle,
\end{align}
with $m_\phi$ the mass of the physical scalar and with $\lambda$ the quartic coupling of the potential of $\phi$, normalized as $\frac{\lambda}{2}|\phi|^4$.
Analogous to \eqref{eq:anomalonboundapp}, we find the following constraint on $g'$
\begin{equation}\label{eq:Apboundyukawa}
g' = \frac{y_\chi}{\sqrt{2}} \frac{m_{A'}}{m_\chi} \lesssim \frac{1}{\sqrt{2}} \frac{m_{A'}}{m_\chi},
\end{equation}
where in the last step, we have taken $y_\chi \lesssim 1$ to avoid the non-perturbative regime. 
In practice, this bound will always be weaker than the bound in \eqref{eq:anomalonboundapp} for the parameter space of interest.

In conclusion, we constructed a UV completion for the axial $A'$ mediator, though it requires a fair amount of additional fields in the ultraviolet to ensure all anomalies are canceled. 
These extra colored particles must be beyond the current LHC limits, which limits the size of $g'$ in the infrared. Moreover, the model predicts the relation
\eqref{eq:anomalouslowE}, resulting in
\begin{equation}
    g_\chi\sim g_n\sim g_p,
    \label{eq:gXnp}
\end{equation}
which is undesirable for direct detection. In the next appendix, we attempt to construct a non-minimal model which is not subject to the parametric constraint in \eqref{eq:gXnp}.


\section{A generalized axial vector model?\label{app:doubleaxialvectormodel}
}

\begin{table}[t]
    \centering
    \begin{tabular}{c|ccccc}
         & SU(3) & SU(2) & $U(1)_Y$ & $U(1)_1$& $U(1)_2$\\\hline
       $Q$  & $\square$ & $\square$ & 1/6 & +1&0\\
       $u$  & $\overline\square$ & $1$ & -2/3 & +1&0\\
       $d$  & $\overline\square$ & $1$ & 1/3 & +1&0\\
       $L$  & 1 & $\square$ & -1/2 & 0&0\\
       $e$  & 1 & 1 & +1 & 0&0\\
       $N$  & 1 & 1 & 0 & +2&0\\
       $H$ & 1 &$\square$ & 1/2 & 0&0\\\hline
       $Q'$  & $\square$ & $\square$ & 1/6 & -1&0\\
       $u'$  & $\overline\square$ & $1$ & -2/3 & -1&0\\
       $d'$  & $\overline\square$ & $1$ & 1/3 & -1&0\\
       $\bar Q'$  & $\overline\square$ & $\square$ & -1/6 & 0&0\\
       $\bar u'$  & $\square$ & $1$ & 2/3 & 0&0\\
       $\bar d'$  & $\square$ & $1$ & -1/3 & 0&0\\
       $ N'$  & 1 & 1 & 0 & -2&0\\
       $\bar N'$  & 1 & 1 & 0 & 0&0\\
       $H_u^\prime$ & 1 & $\square$ & 1/2 &-2 & 0\\
       $H_d^\prime$ & 1 & $\square$ & -1/2 &-2 & 0 \\\hline
       $\chi_1$ &1 &1 & 0 &0&$-1/2$\\
       $\chi_2$ &1 &1 & 0 &0&$-1/2$\\
       $\bar\chi_1$ &1 &1 &0& 0 &$1/2$\\
       $\bar\chi_2$ &1 &1 &0& 0 &$1/2$\\\hline
       $\phi_1$ &1 &1 & 0 &$+1$&0\\
       $\phi_2$ &1 &1 & 0 &0&+1\\
       $\varphi$ &1 &1 & 0 &-1&-1
    \end{tabular}
    \caption{Matter content for an example UV completion of the double axial $A'$ model. The second block and third block are respectively a set of heavy fermions, designed to cancel the gauge anomalies and the dark sector fermion content. The last block contains the scalar sector, in which the $\phi_{1,2}$ generate mass terms for the visible and dark sector fermions, as well as the $A_{1,2}'$. 
    The $\varphi$ field introduces an off-diagonal entry in the $A_{1,2}'$ mass matrix. 
    }
    \label{tab:doubleApmodelcontent}
\end{table}

As shown in \cref{sec:Apdiscussion}, there are no realistic direct detection prospects for sub-GeV dark matter if the dark matter-mediator coupling $g_\chi$ is tied to the mediator-SM coupling $g_p$ for the axial $A'$ model.
It is therefore worthwhile to investigate whether this constraint can be relaxed, such that $g_\chi$ is much larger. In other words, can we construct a model for which one can take $g_\chi \gg g_{p,n}$ in \eqref{eq:UV_vector}?
In this appendix, we argue that it is challenging to write such a model, and one should therefore be wary of assuming $g_{p,n}\ll g_\chi$ without a detailed UV justification. 

\subsection{A second gauge field is needed}

First, one may simply attempt to assign a large charge $|Q_\chi|$ to $\chi_{1,2}$, such that $g_\chi\equiv |Q_\chi| g' \gg g' \sim g_{p,n}$. In \cref{app:AxialVectorModel}, we assigned charges $Q_\chi = -1/2$ so that a single scalar field $\phi$ could generate both the Yukawa terms for SM fields as well as the dark sector masses. There, the charge of the scalar field was fixed by the need to provide mass terms for the anomalon fields in \eqref{eq:anomalonmasses}.
By assigning a large $Q_\chi$, we require an additional scalar field to ensure that the mass terms in \eqref{eq:Apdmmassterms} remain allowed by gauge invariance. Concretely, we now need two fields $\phi_1$ and $\phi_2$ with couplings
\begin{equation}
\mathcal{L} \supset \lambda_Q \phi_1 Q' \bar Q' + \cdots + y_\chi \phi_2 \chi_1\chi_2,  \label{eq:lagneedtwoAp}
\end{equation}
where $\phi_1$ and $\phi_2$ have charge $+1$ and $-2Q_\chi$, respectively, under the $U(1)'$ gauge symmetry.
The model now has two accidental global symmetries which are broken spontaneously by $\langle \phi_1\rangle$ and $\langle \phi_2\rangle$: 
\begin{align}
    &\phi_1 \to e^{-2i\alpha}\phi_1, \;\; Q' \to e^{i\alpha} Q',\;\; \bar Q' \to e^{i\alpha} \bar Q',\;\; \text{etc.}\nonumber\\
    &\phi_2 \to e^{-2i\beta}\phi_2, \;\; \chi_1 \to e^{i\beta} \chi_1,\;\; \chi_2 \to e^{i\beta} \chi_2,\;\; \text{etc.}\label{eq:axialglobal}
\end{align}
One linear combination of the corresponding Goldstone bosons is eaten by the $A'_\mu$, while the other one is exactly massless. 
At one loop, this particle moreover picks up a coupling to photons through the first term in \eqref{eq:lagneedtwoAp}. To avoid violating astrophysical bounds on massless particles with a coupling to photons, one would need to assume that the overlap between the uneaten Goldstone mode and $\phi_1$ is extremely tiny. This would correspond to assuming $\langle \phi_1\rangle \gg Q_\chi \langle \phi_2\rangle$, which is incompatible with our earlier $Q_\chi\gg1$ assumption.

Alternatively, the mass of the Goldstone can be lifted by breaking the symmetries in \eqref{eq:axialglobal} explicitly down to the linear combination corresponding the $U(1)'$ gauge symmetry. 
The lowest dimensional operator that achieves this while respecting gauge invariance is
\begin{equation}
    \mathcal{L} \supset \frac{1}{\Lambda^{2Q_\chi-2}} \phi_1^{2Q_\chi} \phi_2 + \text{h.c.}
\end{equation}
 This operator is extremely irrelevant for $Q_\chi \gg1$ and is therefore not an effective way to lift the mass of the Goldstone.

A straightforward solution is to gauge the second accidental $U(1)'$ symmetry as well, such that the Goldstone mode is eaten.
This means that the visible and dark sectors each have their own axial $U(1)^\prime$ gauge symmetry, whose gauge couplings need not be related to each other.
The interaction between the visible and dark sectors can then be engineered by mixing the two $A'$ fields.
The matter content of this theory is shown in \cref{tab:doubleApmodelcontent}. 
As alluded to before, each $U(1)^\prime$ must be broken by its own scalar field $\phi_{1,2}$, which is also responsible for generating the mass terms of all the fermions in their sector, fully analogous to the discussion in \cref{app:AxialVectorModel}.

\subsection{Heavy mediator limit}

Our primary interest is to investigate whether we can construct a model that is not subject to the $g_p \sim g_\chi$ constraint and reduces to the Hamiltonian \eqref{eq:def_Ap_heavy}. We therefore consider the heavy mediator limit, where $m_{A_1}, m_{A_2}\gg q$.

First, one may consider mixing $A_1'$ and $A_2'$ through a kinetic mixing operator
\begin{equation}
\mathcal{L} \supset\frac{\epsilon}{2} F_1^{\prime\mu\nu}F_{2,\mu\nu}'.
\end{equation}
Because of the derivative couplings in this operator, its effects vanish in the $q\to0$ limit. It therefore does not reproduce \eqref{eq:def_Ap_heavy}.

The mixing between the gauge fields must therefore occur through the mass matrix of the $A'$ fields, rather than their kinetic terms.
To achieve this, we must add a third scalar, $\varphi$, which is charged under both fields. Its vacuum expectation value will mix the $A_{1}'$ and $A_2'$, thus enabling the SM - dark sector interaction that is needed for direct detection.\footnote{This additional scalar once again introduces an unwanted Goldstone boson, however in this instance it can be lifted easily by adding the relevant operator $ \phi_1 \varphi \phi_2$ to the Lagrangian.}
Concretely, the mass matrix for the gauge bosons is
\begin{equation}
    m_A^2 =\left(\!\begin{array}{cc}
      m_1^2   & \delta m^2 \\
       \delta m^2   & m_2^2
    \end{array}\!\right),
\end{equation}
with 
\begin{align}
    m_{1}^2 =& 2 g_{1}^2 (|\langle \phi_{1} \rangle|^2+|\langle \varphi \rangle|^2)\nonumber\\
    m_{2}^2 =& 2 g_{2}^2 (|\langle \phi_{2} \rangle|^2+|\langle \varphi \rangle|^2)\nonumber\\
    \delta m^2 =& 2 g_1 g_2  |\langle \varphi \rangle|^2, \label{eq:axialm}
\end{align}
Since we will assume $g_1 \ll g_2\sim 1$, it follows that $\delta m \ll m_2$ regardless of relative size of the three vacuum expectation values.

First, let us consider the case with $m_1<m_2$.
Denoting the lightest and heaviest mass eigenstates with $A^\mu_\ell$ and $A^\mu_h$, respectively, we find
\begin{align}
m^2_{A_\ell} &\approx m_1^2 - \frac{\delta m^4}{m_2^2-m_1^2}\\
m^2_{A_h} &\approx m_2^2 + \frac{\delta m^4}{m_2^2-m_1^2},
\end{align}
where we expanded in $\delta m^2\ll m_2^2-m_1^2$.
The mass eigenstates couple to the visible and dark sector currents through 
\begin{align}
    \mathcal{L}\supset &A_\ell^\mu \left(g_1 J_{1,\mu} - g_2 \frac{\delta m^2}{m_2^2-m_1^2}J_{2,\mu}\right)\\
    &+A_h^\mu \left(g_2 J_{2,\mu} + g_1 \frac{\delta m^2}{m_2^2-m_1^2}J_{1,\mu}\right),
\end{align}
where $J_{1,\mu}$ and $J_{2,\mu}$ are the SM and dark matter axial currents
\begin{align}
    J_{1}^\mu &\equiv \bar q \gamma^\mu \gamma_5 q\\
    J_{2}^\mu &\equiv -\frac{1}{2}\bar \chi \gamma^\mu \gamma_5 \chi.
\end{align}
The leading contribution to the direct detection scattering rate will be through the exchange of the $A_\ell$.
The Hamiltonian for direct detection is that in \eqref{eq:def_Ap_heavy} with
\begin{align}
g_{n,p} &\sim g_1\\
g_\chi & =  \frac{1}{2} \frac{\delta m^2}{m_2^2-m_1^2} g_2\\
m_{A'}& = m_{A_\ell},
\end{align}
where we have used the $\sim$ since e.g., the neutron/proton coupling will also contain a form factor (see \ref{eq:formFactorAxialVec}). 
Plugging in \eqref{eq:axialm}, we find that 
\begin{equation}\label{eq:thismodelsucks}
    |g_\chi| \leq  \frac{|g_{p,n}|}{2},
\end{equation}
which is incompatible with the $|g_\chi| \gg |g_{p,n}|$ hierarchy we seek.

Alternatively, we could have chosen $m_1>m_2$ by taking $\langle \phi_1\rangle \gg \langle \phi_2\rangle, \langle \varphi\rangle$. The $A_\ell$ now predominantly aligns with the dark sector $U(1)'$ ($A_2^\mu$) and we find
\begin{align}
m^2_{A_\ell} &\approx m_2^2 - \frac{\delta m^4}{m_1^2-m_2^2}\\
m^2_{A_h} &\approx m_1^2 + \frac{\delta m^4}{m_1^2-m_2^2},
\end{align}
and
\begin{align}
    \mathcal{L}\supset &A_\ell^\mu \left(-g_2 J_{2,\mu} + g_1 \frac{\delta m^2}{m_1^2-m_2^2}J_{1,\mu}\right)\\
    &+A_h^\mu \left(g_1 J_{1,\mu} + g_2 \frac{\delta m^2}{m_1^2-m_2^2}J_{2,\mu}\right).
\end{align}
To avoid suppressing the mixing angle further, we need to take $m_1\sim m_2$, since we already have the hierarchy $\delta m\ll m_2$ from \eqref{eq:axialm}. 
There is therefore only a small hierarchy between the mass eigenstates \mbox{($m_{A_\ell}\lesssim m_{A_h}$)}.
All the bounds in \cref{sec:modelAp} moreover apply to the $A_h$, such that we must still impose $g_1\ll 1$. 

Again plugging in \eqref{eq:axialm}, we now find that the couplings of the $A_\ell$ can be identified with \eqref{eq:def_Ap_heavy} through
\begin{align}
g_{n,p} &\sim \frac{\delta m^2}{m_1^2-m_2^2} g_1\lesssim \frac{g_1^2}{g_2}\\
g_\chi & = g_2.
\end{align}
This realizes the desired hierarchy $g_{p,n}\ll g_\chi$. However, the scattering amplitude for direct detection is proportional to their product
\begin{equation}
  \mathcal{M}_{p\chi \to p \chi} \propto  g_{p,n}g_\chi \lesssim g_1^2,
\end{equation}
where $g_1$ is subject to the bounds in \cref{fig:Aprime_constraints} because of the presence of the $A_h$, with $m_{A_h}\sim m_{A_\ell}$. 
In other words, the effective suppression factor is again the same as what we have found before.


As a last resort, one may attempt to fine tune the vacuum expectation values such that $\delta m^2 \sim m_2^2 - m_1^2$ even though  $\delta m \ll m_2, m_1$. It is now convenient to introduce the notation
\begin{align}
\bar m^2 \equiv \frac{1}{2} (m_1^2+m^2_2)\\
\Delta m^2 \equiv \frac{1}{2} (m_1^2-m^2_2).
\end{align}
The physical masses are now
\begin{align}
m^2_{A_\ell} &= \bar m^2 (1- \sqrt{\delta m^4 + \Delta m^4}/\bar m^2)\\
m^2_{A_h} &= \bar m^2 (1+ \sqrt{\delta m^4 + \Delta m^4}/\bar m^2).
\end{align}
The couplings are of the form
\begin{align}
    \mathcal{L}\supset &A_\ell^\mu \left( \cos\theta g_1 J_{SM,\mu} +  \sin\theta g_2 J_{\chi,\mu}\right)\nonumber\\
    &+A_h^\mu \left(- \sin\theta g_1 J_{SM,\mu} + \cos \theta g_2 J_{\chi,\mu}\right),\label{eq:nowwearereallydead}
\end{align}
with $\theta$ the mixing angle, as determined by the values of $\delta m$ and $\Delta m$. 
One can take it to be $\mathcal{O}(1)$ by tuning $\delta m \approx \Delta m$.
From \eqref{eq:nowwearereallydead}, we now see that $A_\ell^\mu$ and $A_h^\mu$ will destructively interfere, with the cancellation being exact in the limit where $m_{A_\ell} \approx m_{A_h}$.
At the level of the amplitude, if we expand in $\Delta m, \delta m \ll \bar m$, this cancellation is only lifted by corrections that are suppressed by $\sim \sqrt{\delta m^4 +\Delta m^4}/\bar m^2\ll 1$, thus recovering the same suppression factor we found before.

In summary, we find it very difficult to build a model of a heavy axial vector mediator which can accommodate a parametrically larger direct detection rate than what we presented in \cref{sec:modelAp}. We currently are not aware of a mechanism which circumvents the argument laid out in this section.

\subsection{A light mediator model}

For completeness, we comment briefly on the light mediator regime of the $U(1)' \times U(1)'$ model ($m_{A_h},m_{A_\ell}\ll q$). In this case, we can couple the $A_1'$ and $A_2'$ through kinetic mixing. 
After diagonalizing to the mass basis and taking the non-relativistic limit, we find the effective Hamiltonian
\begin{equation}
\label{eq:light_u1u1}
    \mathcal{H}^{\rm light}_{A_1^\prime\otimes A_2^\prime} = 4 g_p g_\chi \epsilon \left [ \frac{(\mathbf{q}\cdot \mathbf{J}_\chi)(\mathbf{q}\cdot \mathbf{S}_p)}{|\mathbf{q}|^4} - \frac{\mathbf{J}_\chi\cdot\mathbf{S}_p}{|\mathbf{q}|^2} \right ] e^{i\mathbf{q}\cdot \mathbf{r}},
\end{equation}
with 
\begin{equation}
g_p \approx 0.22 g_1 \quad \mathrm{and} \quad g_\chi=\frac{1}{2}g_2    
\end{equation}
after matching the quark current to the proton (see \cref{sec:modelAp}). 
The scattering rate is then given by
\begin{multline}
    R^{\rm light}_{A_1^\prime\otimes A_2^\prime} \simeq \frac{\bar{\sigma}^{\rm light}_{A_1^\prime\otimes A_2^\prime}}{\sum_dm_d} \frac{\rho_\chi}{m_\chi} \frac{\pi}{3} \frac{q_0^4}{v_0^4 \mu_{p\chi}^6} (J_\chi +1)J_\chi \\ \int d^3v \, f(\mathbf{v}) \int \frac{d^3 q}{(2\pi)^3} |F(\mathbf{q})|^2 S(\mathbf{q},\omega),
\end{multline}
with the reference cross section
\begin{equation}
    \bar{\sigma}^{\rm light}_{A_1^\prime\otimes A_2^\prime} \simeq \frac{64 g_p^2 g_2^2 \epsilon^2 v_0^4 \mu_\chi^6}{3\pi q_0^8}.
\end{equation}
This model realizes the desired $g_{p}\ll g_\chi$ hierarchy and is therefore appears promising. 
However, we assumed $m_{A_h},m_{A_\ell}\ll q \sim v_0 m_\chi$ to derive \eqref{eq:light_u1u1}, which implies extremely strong bounds on $g_p$ (see \cref{fig:Aprime_constraints}).
When accounting for these existing bounds, we find no open parameter space for direct detection, even when setting $g_\chi=1$. 

\section{Implementation in DarkELF\label{app:darkelf}}


We have implemented in \textbf{DarkELF} the ability to calculate rates for four different spin dependent models. The rates  used for the models are slightly modified versions from those given in the bulk of the paper, so that an energy threshold can be applied. Specifically, introducing a factor of $\int d \omega \delta(\omega - \bfq \cdot \bfv + q^2/(2 m_\chi) )$ and performing the integral over velocity $v$ first gives: 
\begin{align}
    R_\phi &= \frac{\bar\sigma_\phi}{\sum_d m_d} \frac{\rho_\chi}{m_\chi}  \frac{1}{3} \frac{  m_p^2}{v_0^2 \mu_{\chi p}^4} \int\limits_{\omega_{th}}^{\omega_+} \!\! d\omega   \int\limits_{q_-}^{q_+} \!\! \frac{dq}{2\pi}  \, \nonumber\\ 
    &\hspace{1mm}\times |F_\phi(q)|^2 \frac{|q|^3}{m_p^2} S(q,\omega) \, \eta(v_{min}(q,\omega))\\
    R_a &=   \frac{\bar\sigma_a}{\sum_d m_d}\frac{\rho_\chi}{m_\chi} \frac{1}{8} \frac{ m_p^2 m_\chi^2}{v_0^4 \mu_{p\chi}^6}  \int\limits_{\omega_{th}}^{\omega_+} \!\! d\omega   \int\limits_{q_-}^{q_+} \!\!\frac{dq}{2\pi} \, \nonumber\\ 
    &\hspace{1mm}\times |F_a(q)|^2 \frac{|q|^5}{m_p^2 m_\chi^2} S(q,\omega) \, \eta(v_{min}(q,\omega))\\
    R_{A'}^{\rm heavy} &=  \frac{\bar\sigma_{A'}^{\rm heavy}}{\sum_d m_d} \frac{\rho_\chi}{m_\chi}   \frac{2}{3}  \frac{1}{\mu_{p\chi}^2} \int\limits_{\omega_{th}}^{\omega_+} \!\! d\omega    \int\limits_{q_-}^{q_+} \!\!\frac{dq}{2\pi} \, \nonumber\\ 
    &\hspace{1mm}\times |F_{A'}(q)|^2  |q|S(q,\omega) \, \eta(v_{min}(q,\omega))\\
    R^{\rm light}_{A_1^\prime\otimes A_2^\prime} &\simeq \frac{\bar{\sigma}^{\rm light}_{A_1^\prime\otimes A_2^\prime}}{\sum_dm_d} \frac{\rho_\chi}{m_\chi} \frac{1}{8} \frac{q_0^4}{v_0^4 \mu_{p\chi}^6} \int\limits_{\omega_{th}}^{\omega_+} \!\! d\omega    \int\limits_{q_-}^{q_+} \!\!\frac{dq}{2\pi} \, \nonumber\\ 
    &\hspace{1mm}\times |F_{A'}(q)|^2  |q|S(q,\omega) \, \eta(v_{min}(q,\omega)),
\end{align}
where the integration limits are
\begin{align}
    q_\pm &\equiv m_\chi v_{max} \left( 1 \pm \sqrt{1 - \frac{2\omega_{th}}{m_\chi v_{max}^2}} \, \right) \\
    \omega_+  &\equiv \frac{1}{2}m_\chi v_{max}^2,
\end{align}\\
and $v_{max} = v_{esc} + v_e$ which is the maximum DM speed in the lab frame. $\eta$ is defined in terms of the DM velocity distribution as:
\begin{equation}
    \eta(v_{min}) = \int d^3v \frac{f(v)}{v}\Theta(v-v_{min}),
\end{equation}
where $v_{min} = \frac{q}{2m_\chi} + \frac{\omega}{q}$. The structure factor $S(q, \omega)$ uses the calculation of the correlation function $C_{\ell d}$, \eqref{eq:Cldresult}, which is described in \cite{Campbell-Deem:2019hdx}. For \mbox{$q > 2 \sqrt{ 2 A_d m_p \bar \omega_d}$}, this calculation of $C_{\ell d}$ switches over to using the impulse approximation, which gives a simple approximation for the correlation function when many phonons can be produced.

The code also contains the free nuclear recoil limit of these expressions, relevant for heavier DM masses. This amounts to taking the limit $q, \omega \rightarrow \infty$, which modifies $C_{\ell d}$ and the structure factor in the following way:
\begin{align}
    \lim_{q, \omega \to \infty} C_{\ell, d} &= 2\pi \, \delta \left(\omega - \frac{q^2}{2m_d} \right) \\
    S^{FR}(q, \omega) &= 2\pi \sum_d \overline{\lambda_d^2 J_d(J_d + 1)} \, \delta \left(\omega - \frac{q^2}{2m_d} \right).
\end{align}
In this limit, the rate expressions also simplify dramatically: 
\begin{align}
    \frac{dR_\phi}{d\omega} &= \frac{2}{3}\frac{\rho_\chi}{m_\chi} \frac{N_{uc} \bar\sigma_\phi}{v_0^2 \mu_{\chi p}^4}  \omega \nonumber\\ 
    &\hspace{-2mm}\times \sum_d \overline{\lambda_d^2 J_d(J_d + 1)} \, m_d^2 \, |F_{\phi}(q)|^2\, \eta(v_{min})  \\
    \frac{dR_a}{d\omega} &=  \frac{1}{2} \frac{\rho_\chi}{m_\chi}  \frac{N_{uc} \bar\sigma_a}{v_0^4 \mu_{p\chi}^6} \omega^2 \nonumber\\ 
    &\hspace{-2mm}\times \sum_d \overline{\lambda_d^2 J_d(J_d + 1)} \, m_d^3 \, |F_{a}(q)|^2 \, \eta(v_{min})  \\
    \frac{dR_{A'}^{\rm heavy}}{d\omega} &= \frac{2}{3}\frac{\rho_\chi}{m_\chi} \frac{N_{uc} \bar\sigma_{A'}}{\mu_{p\chi}^2} \nonumber\\
    &\hspace{-2mm}\times \sum_d \overline{\lambda_d^2 J_d(J_d + 1)} \, m_d \, |F_{A'}(q)|^2\, \eta(v_{min})  \label{eq:dR_domega_A'}\\
    \frac{dR^{\rm light}_{A_1^\prime\otimes A_2^\prime}}{d\omega} &\simeq \frac{1}{8} \frac{\rho_\chi}{m_\chi} \frac{N_{uc} \bar{\sigma}^{\rm light}_{A_1^\prime\otimes A_2^\prime}}{\mu_{p\chi}^6} m_\chi^4  \nonumber\\ 
    &\hspace{-2mm}\times \sum_d \overline{\lambda_d^2 J_d(J_d + 1)} \, m_d \, |F_{A'}(q)|^2 \, \eta(v_{min}),
\end{align}
where $N_{uc} = 1/\sum_d m_d$ is the number of unit cells per unit target mass and $q = \sqrt{2m_d\omega}$. 
For \mbox{$\omega > 1$ eV}, the code switches over to using this limiting expression directly which eliminates some previous numerical noise. In order to only plot the nuclear recoil limit for the spin dependent case, one can change the flag \texttt{nuclear\_recoil} to \texttt{True} in either \texttt{sigma\_multiphonons\_SD()} or \texttt{R\_multiphonons\_SD()}. Note that the expression for $dR^{\rm heavy}_{A'}/d\omega$ in \eqref{eq:dR_domega_A'} can be rewritten in a form that is used more widely in the literature. Rewriting the factor of $ N_{uc} m_d = m_d/ \sum_d m_d$ as the mass fraction of each nucleus and replacing $\lambda_d = \langle S_p \rangle / J_d$, then  
\begin{align}
    \frac{dR_{A'}^{\rm heavy}}{d\omega} &= \frac{2 \rho_\chi}{m_\chi} \frac{\bar\sigma_{A'}}{\mu_{p\chi}^2} \nonumber\\
    &\hspace{-15mm}\times \sum_d  \frac{m_d}{\sum_{d'} m_{d'}} \, \frac{\langle S_p \rangle^2 (J_d + 1)}{3 J_d} \, |F_{A'}(q)|^2\, \eta(v_{min}),
\end{align} 
as in \cite{CRESST:2022dtl}.

 \begin{table*}[ht]
     \centering
     \begin{tabularx}{\textwidth}{XXl}
        Function  &  Location & Description \\
        \hline \hline
        \texttt{sigma\_multiphonons\_SD(threshold)}  & \texttt{multiphonon\_spin\_dependent.py} & \begin{tabular}{l}
            Spin dependent nucleon cross section to \\
            produce 3 events/kg-yr  
        \end{tabular} \\
        \texttt{R\_multiphonons\_SD(threshold)} & \texttt{multiphonon\_spin\_dependent.py} & \begin{tabular}{l} Total phonon rate for three possible spin \\
        dependent operators in units of 1/kg-yr \end{tabular}\\
        \texttt{update\_params()} & \texttt{init.py} & \begin{tabular}{l}
            Function to change settings in the DM\\
            model, including DM mass, SD operator, etc.  
        \end{tabular} \\
     \end{tabularx}
     \caption{List of public functions in DarkELF related to both spin dependent multiphonon excitations from DM scattering. Only mandatory arguments are shown; for optional arguments and flags, see text and the documentation in repository. \texttt{threshold} denotes the energy threshold of the experiment.} 
     \label{tab:public_functions}
 \end{table*}
    
By default, the code will return the scattering rate for the $A'$ mediator. In order to change this, use \texttt{update\_params(SD\_op = new\_op)} where \texttt{new\_op} can be selected from the list \texttt{\{ "A'", "phi", "a", "double A'"\}}. The other parameter that can be changed via \texttt{update\_params()} is the value of the ratio between $g_p$ and $g_n$. By default, the cross sections $\bar \sigma$ refer to spin-dependent scattering off protons while the ratio $g_n/g_p$ is incorporated into the factor $\lambda_d$ as defined in \eqref{eq:deflambda}. For $\phi$ and $a$, this ratio is fixed to be the value predicted for the UV completion described in the paper. For $A'$, we set $g_n / g_p $ = 1. These values can be changed via the keyword argument: \texttt{update\_params(gp\_gn\_ratio\_val = new\_val, set\_gp\_gn\_ratio\_val = True)}. To evaluate cross sections in terms of scattering on neutrons instead, one can also change the keyword argument \texttt{gp\_gn\_ratio} from \texttt{"g\_n/g\_p"} to be \texttt{"g\_p/g\_n"} and specify a ratio for $g_p/g_n$. All public functions are summarized in \cref{tab:public_functions}.

The spin dependent code requires several specific parameters in the yaml file for each material. This has so far been implemented for GaAs and Al$_2$O$_3$, but the user can add additional materials as follows: The \texttt{unitcell} dictionary contains all necessary parameters in a convenient form. The dictionary should have an entry for each atom in the unit cell which will form its own dictionary, containing the atomic number, multiplicity in the unit cell and information about the atom's isotopes. For example, the `Ga' dictionary takes the form:
\begin{verbatim}
"Ga":{"A":69, "mult":1, "isotopes":[{"A":69., 
        "frac":0.601, "atomic_spin":1.5, 
        "f_p":0.0750967, "f_n":0.0}, {"A":71., 
        "frac":0.399, "atomic_spin":1.5, 
        "f_p":0.154422, "f_n":0.0}]}
\end{verbatim}
All isotopes should be included such that the isotope fractions add up to 1. These fractions can either be the natural abundances or the specific abundances of the particular material of interest. $f_p$ and $f_n$ can be obtained via the Odd-Group Model or shell model calculations, as preferred (see \cref{table:response_functions}).

\bibliography{spin_dep}

\end{document}